\pdfoutput=1
\documentclass[11pt,twoside,a4paper,cmspaper,final,collab]{cms-tdr}
\def\svnVersion{efd6216}\def\svnDate{2023/07/13}\def\cmsCernNoTag{CERN-EP-2022-158}\def\cmsCernDate{\today}\def\cmsMessage{Published in the Journal of High Energy Physics as \href{http://dx.doi.org/10.1007/JHEP07(2023)046}{\doi{10.1007/JHEP07(2023)046}.}}
\begin{document}\cmsNoteHeader{TOP-21-010}

\newlength\cmsTabSkip\setlength{\cmsTabSkip}{1ex}

\newcommand{\tW}{\ensuremath{\PQt\PW}\xspace}
\newcommand{\emu}{\ensuremath{\Pepm\PGmmp}\xspace}
\newcommand{\PHP}{{\POWHEG} + {\PYTHIA}8\xspace}
\newcommand{\PHH}{{\POWHEG} + {\HERWIG}7\xspace}
\newcommand{\pz}{\ensuremath{p_z}\xspace}
\newcommand{\dy}{Drell--Yan\xspace}
\newcommand{\pp}{\ensuremath{\Pp\Pp}\xspace}
\newcommand{\roots}{\ensuremath{\sqrt{s}}\xspace}
\newcommand{\rootsthirteen}{\ensuremath{\roots=13\TeV}\xspace}
\newcommand{\rootseight}{\ensuremath{\roots=8\TeV}\xspace}
\newcommand{\tWxsec}{\ensuremath{\sigma_{\tW}}\xspace}
\newcommand{\tWxsecSM}{\ensuremath{\tWxsec^{\text{SM}}}\xspace}
\newcommand{\tWxsecobs}{\ensuremath{79.2\pm0.9\stat\,^{+7.7}_{-8.0}\syst\pm1.2\lum\unit{pb}}\xspace}
\newcommand{\lumiwounc}{\ensuremath{138\fbinv}\xspace}
\newcommand{\pzvar}{\ensuremath{\pz(\Pepm, \PGmmp, \PQj)}\xspace}
\newcommand{\invmassvar}{\ensuremath{m(\Pepm, \PGmmp, \PQj)}\xspace}
\newcommand{\transmassvar}{\ensuremath{\mT(\Pepm, \PGmmp, \PQj, \ptvecmiss)}\xspace}
\newcommand{\deltaPhiVar}{\ensuremath{\Delta\varphi(\Pepm, \PGmmp)}\xspace}
\newcommand{\mtop}{\ensuremath{m_{\PQt}}\xspace}
\newcommand{\PQj}{\ensuremath{\HepParticle{j}{}{}}\xspace}
\newcommand{\njnb}[2]{\ensuremath{#1\PQj#2\PQb}\xspace}
\newcommand{\VV}{\ensuremath{\PV\PV}\xspace}
\newcommand{\ttV}{\ensuremath{\ttbar\PV}\xspace}
\newcommand{\abseta}{\ensuremath{\abs{\eta}}\xspace}
\newcommand{\vecn}{\ensuremath{\vec{n}}\xspace}
\newcommand{\thetaj}{\ensuremath{\theta_j}\xspace}
\newcommand{\vectheta}{\ensuremath{\vec{\theta}}\xspace}
\newcommand{\vecpTi}{\ensuremath{\vec{p}_{\mathrm{T},i}}\xspace}
\newcommand{\muR}{\ensuremath{\mu_{\mathrm{R}}}\xspace}
\newcommand{\muF}{\ensuremath{\mu_{\mathrm{F}}}\xspace}
\newcommand{\hattheta}{\ensuremath{\hat{\theta}}\xspace}
\newcommand{\thetazero}{\ensuremath{\theta_0}\xspace}
\newcommand{\dtheta}{\ensuremath{\Delta\theta}\xspace}
\newcommand{\pull}{\ensuremath{(\hattheta-\thetazero)/\dtheta}\xspace}
\newcommand{\nonWZ}[1][n]{\ensuremath{\text{#1on-}\PW\!/\PZ}\xspace}
\newcommand{\dhatmu}{\ensuremath{\Delta\hat{\mu}}\xspace}
\newcommand{\etaphi}{\ensuremath{\eta\text{-}\varphi}\xspace}

\cmsNoteHeader{TOP-21-010}
\title{Measurement of inclusive and differential cross sections for single top quark production in association with a \PW boson in proton-proton collisions at \texorpdfstring{\rootsthirteen}{sqrt(s) = 13 TeV}}

\date{\today}

\abstract{Measurements of the inclusive and normalised differential cross sections are presented for the production of single top quarks in association with a \PW boson in proton-proton collisions at a centre-of-mass energy of 13\TeV. The data used were recorded with the CMS detector at the LHC during 2016--2018, and correspond to an integrated luminosity of \lumiwounc. Events containing one electron and one muon in the final state are analysed. For the inclusive measurement, a multivariate discriminant, exploiting the kinematic properties of the events is used to separate the signal from the dominant \ttbar background. A cross section of \tWxsecobs is obtained, consistent with the predictions of the standard model. For the differential measurements, a fiducial region is defined according to the detector acceptance, and the requirement of exactly one jet coming from the fragmentation of a bottom quark. The resulting distributions are unfolded to particle level and agree with the predictions at next-to-leading order in perturbative quantum chromodynamics.}

\hypersetup{
    pdfauthor={CMS Collaboration},
    pdftitle={Measurement of inclusive and differential cross sections for single top quark production in association with a W boson in proton-proton collisions at sqrt(s) = 13 TeV},
    pdfsubject={CMS},
    pdfkeywords={CMS, physics, top quark}
}

\maketitle
\section{Introduction}
The electroweak production of single top quarks was first observed by the \DZERO~\cite{Abazov:2009ii} and CDF~\cite{Aaltonen:2009jj} Collaborations at the Fermilab Tevatron. At leading order, single top quark production proceeds mainly via three processes: the $t$-channel exchange of a virtual \PW boson, $s$-channel production, and the associated production of a top quark and a \PW boson (\tW). The last channel, which was negligible at the Tevatron, represents a significant contribution to single top quark production at the CERN LHC. It is a very interesting production mechanism because of its interference with top quark pair (\ttbar) production~\cite{Frixione:2008yi,Belyaev:1998dn,White:2009yt}, its sensitivity to beyond the standard model (SM) physics~\cite{Tait:2000sh,Cao:2007ea,Barger:2009ky,ATLAS:2017hoo,CMS:2019zzl,Pinna:2017tay}, and its role as a background in several other analyses.

\begin{figure}[htbp!]
    \centering
    \includegraphics[width=0.35\textwidth]{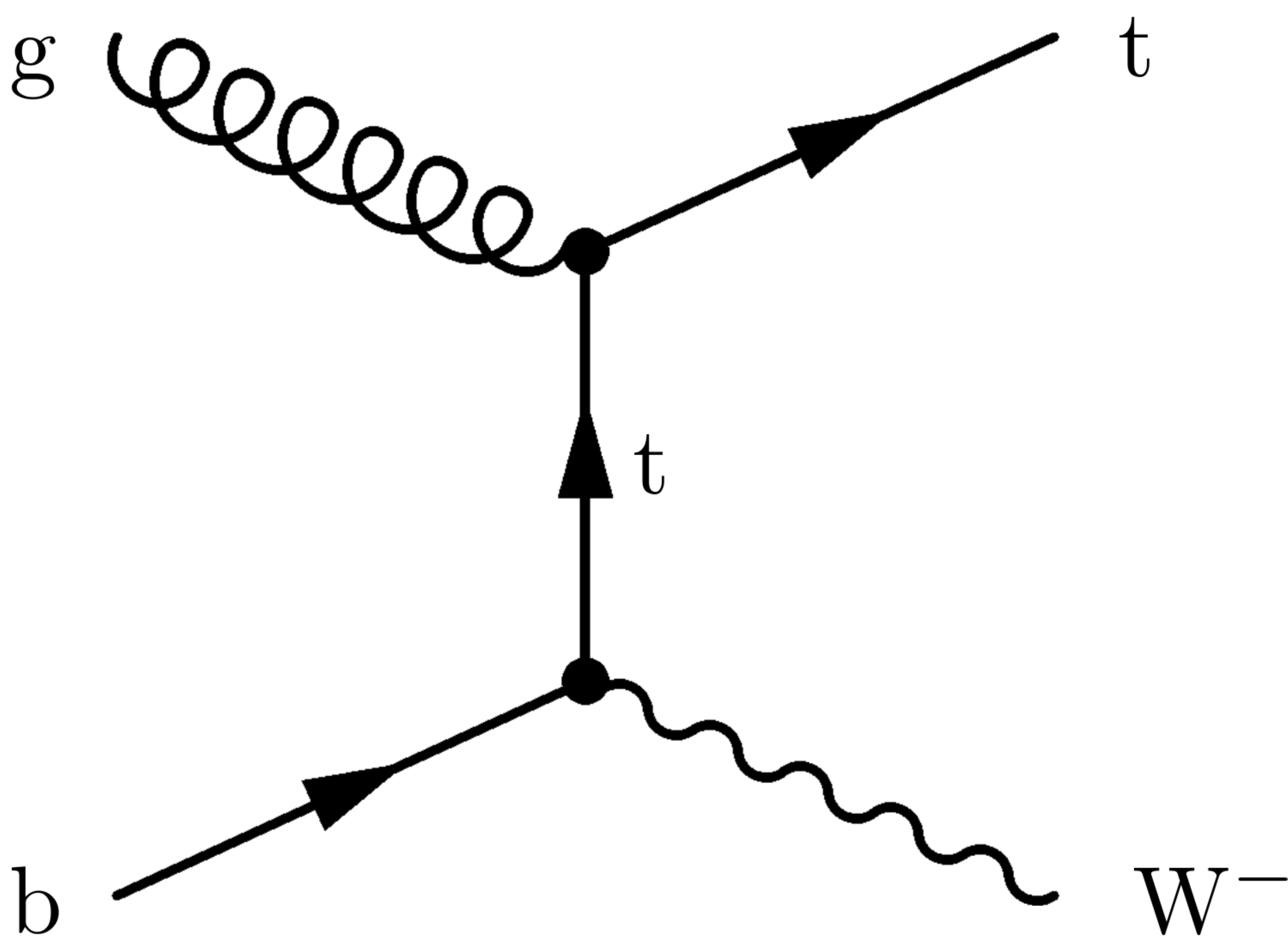}%
    \hspace*{1.5cm}%
    \includegraphics[width=0.35\textwidth]{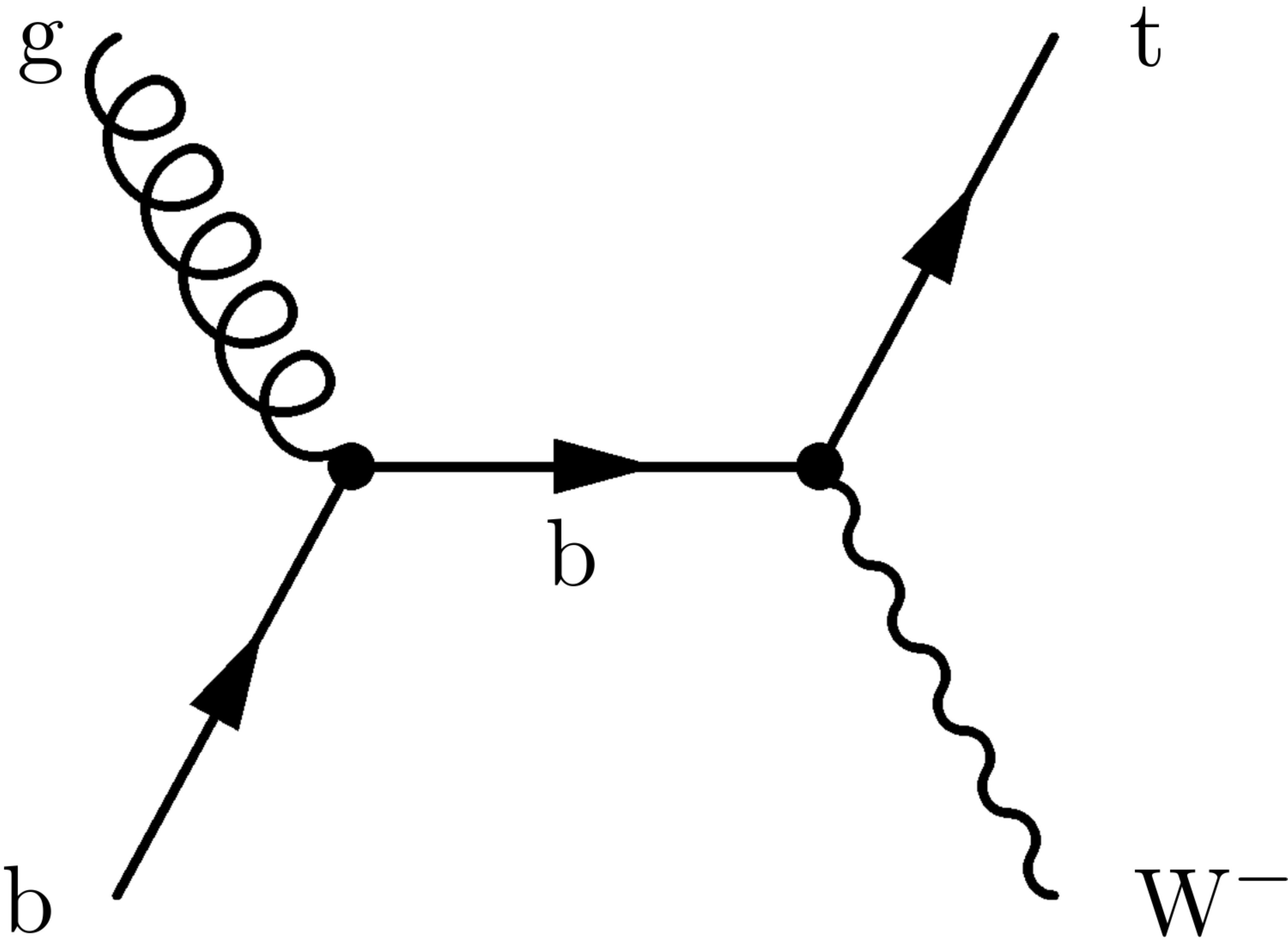}
    \caption{Leading-order Feynman diagrams for single top quark production in the \tW mode. The charge-conjugate modes are implicitly included.}
    \label{fig:diags}
\end{figure}

The definition of \tW production at next-to-leading order (NLO) in perturbative quantum chromodynamics (QCD) shares final states with \ttbar production~\cite{Frixione:2008yi,Belyaev:1998dn,White:2009yt}. The cross section for \tW production has been computed at approximate next-to-NLO order (NNLO) in QCD. Assuming a top quark mass (\mtop) of 172.5\GeV, the theoretical NNLO cross section for \tW production in proton-proton (\pp) collisions at \rootsthirteen is $\tWxsecSM = 71.7 \pm 1.8\,\text{(scale)} \pm 3.4\,\text{(PDF)}\unit{pb}$~\cite{Kidonakis:2015nna}, where the first uncertainty corresponds to the renormalisation (\muR) and factorisation (\muF) scale variations, and the second to using different parton distribution function (PDF) sets. The leading-order (LO) Feynman diagrams for \tW production are shown in Fig.~\ref{fig:diags}.

The \tW channel was not accessible at the Tevatron due to its small cross section in $\Pp\PAp$ collisions at $\roots=1.96\TeV$. At the LHC, however, evidence for this process in 7\TeV collision data was presented by ATLAS~\cite{Aad:2012xca} and CMS~\cite{Chatrchyan:2012zca}. At $\roots=8\TeV$, measurements by CMS~\cite{Chatrchyan:2014tua} and ATLAS~\cite{Aad:2015eto} were in good agreement with theoretical predictions.

At \rootsthirteen the inclusive \tW production cross section has been measured by both ATLAS~\cite{Aaboud:2016lpj} and CMS~\cite{Sirunyan:2018lcp} using data recorded during 2016. Both measurements employed dileptonic (\emu for CMS and \EE, \MM, and \emu for ATLAS) channels. The measurement of the differential \tW production cross section is particularly challenging due to the overwhelming background from \ttbar events, with 80\%~\cite{Aaboud:2016lpj} in the most \tW-enriched category. The ATLAS experiment performed the first measurement of the \tW production differential cross section~\cite{Aaboud:2017qyi}, and a study~\cite{ATLAS:2018ivx} of the $\PW\PW\PQb\PQb$ final state (that includes \tW and \ttbar). Studies in the semileptonic channel have also been done recently at \rootsthirteen by CMS~\cite{CMS:2021vqm} and at \rootseight by ATLAS~\cite{ATLAS:2020cwj}.

This paper reports a measurement of inclusive and normalised differential \tW production cross sections at \rootsthirteen with dilepton final states (\emu), using data collected with the CMS detector during 2016--2018, corresponding to an integrated luminosity of \lumiwounc. The paper is structured as follows. A summary  of the data and Monte Carlo (MC) samples used is provided in Section~\ref{sec:montecarlo}. The object and event selection criteria are discussed in Section~\ref{sec:eventselection}. The signal extraction for the inclusive and differential measurements are detailed in Sections~\ref{sec:inclusive} and~\ref{sec:differential}, respectively. The systematic uncertainties are discussed in Section~\ref{sec:uncs}. The inclusive and differential results are presented in Section~\ref{sec:results}. Finally, a summary of both measurements is given in Section~\ref{sec:summary}. Tabulated results are provided in the HEPData record for this analysis~\cite{hepdata}.

\section{The CMS detector and MC simulation}
\label{sec:montecarlo}
The central feature of the CMS apparatus is a superconducting solenoid of 6\unit{m} internal diameter, providing a magnetic field of 3.8\unit{T}. Within the solenoid volume are a silicon pixel and strip tracker, a lead tungstate crystal electromagnetic calorimeter, and a brass and scintillator hadron calorimeter, each composed of a barrel and two endcap sections. Forward calorimeters extend the pseudorapidity ($\eta$) coverage provided by the barrel and endcap detectors. Muons are detected in gas-ionisation chambers embedded in the steel flux-return yoke outside the solenoid. A more detailed description of the CMS detector, together with a definition of the coordinate system used and the relevant kinematic variables, can be found in Ref.~\cite{Chatrchyan:2008zzk}. Events of interest are selected using a two-tiered trigger system. The first level (L1), composed of custom hardware processors, uses information from the calorimeters and muon detectors to select events at a rate of around 100\unit{kHz} within a fixed latency of about 4\mus~\cite{CMS:2020cmk}. The second level, known as the high-level trigger, consists of a farm of processors running a version of the full event reconstruction software optimized for fast processing, and reduces the event rate to around 1\unit{kHz} before data storage~\cite{CMS:2016ngn}.

\begin{figure}[htbp!]
    \centering
    \includegraphics[width=0.32\textwidth]{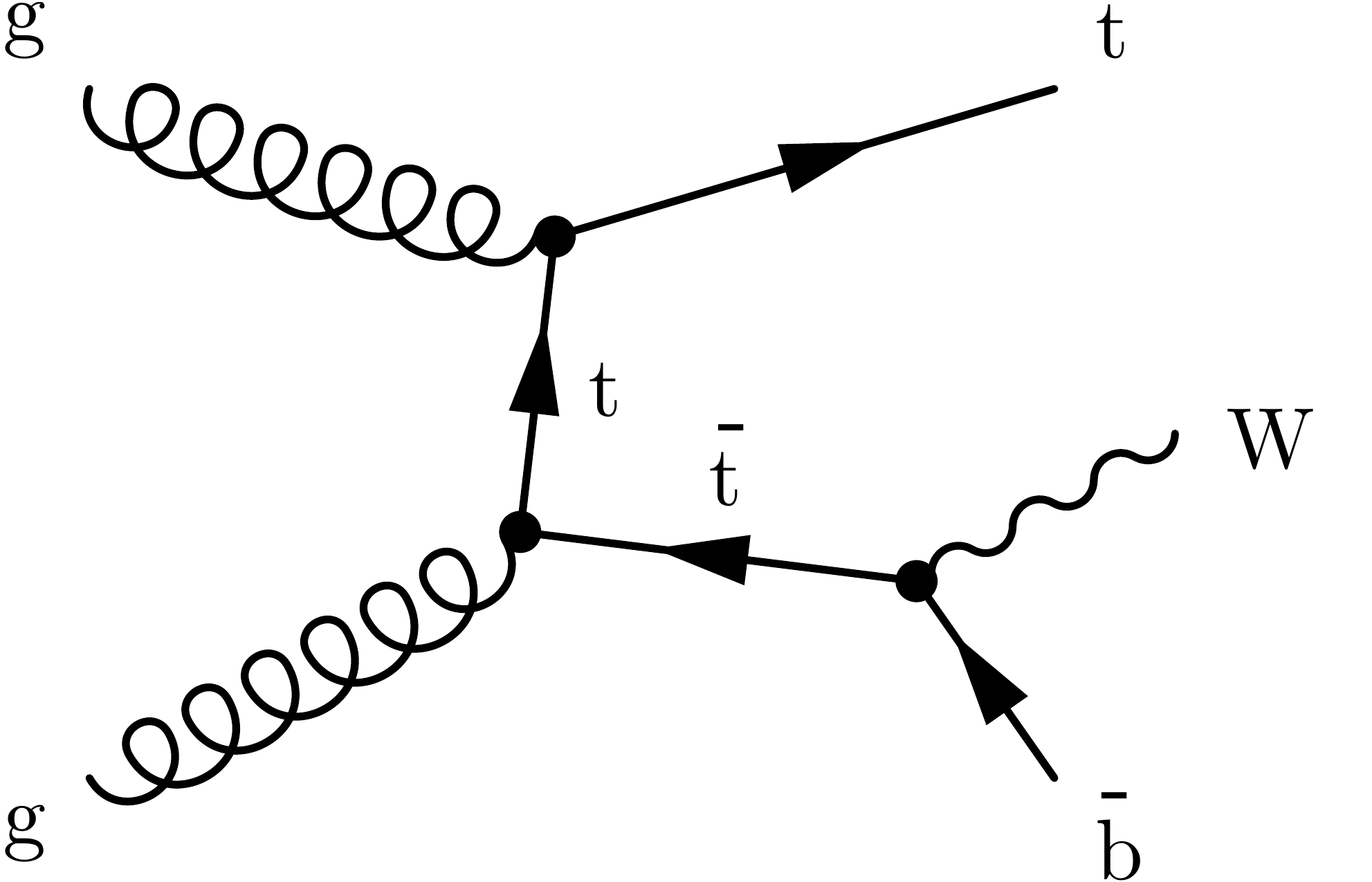}%
    \hfill%
    \includegraphics[width=0.32\textwidth]{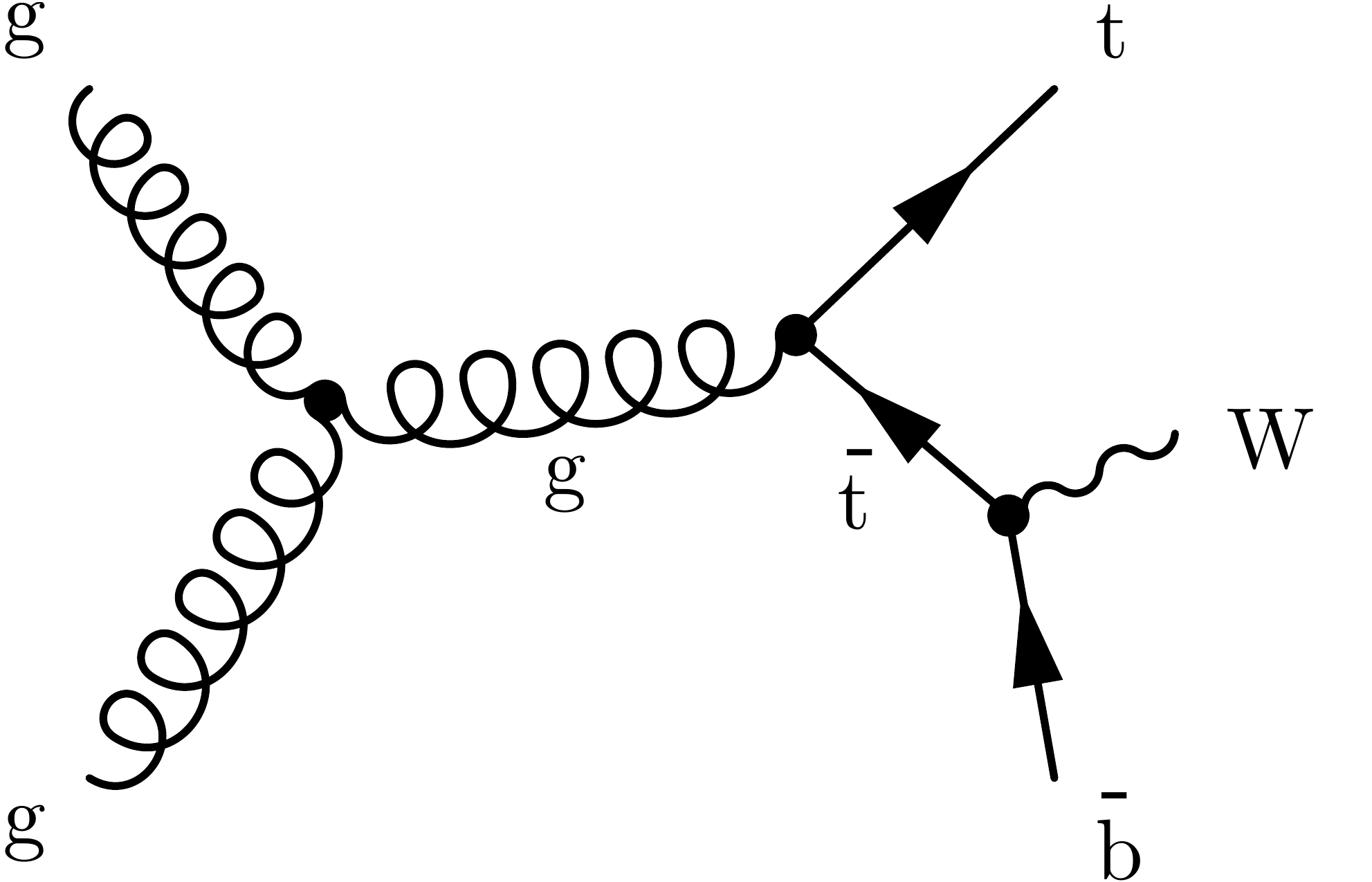}%
    \hfill%
    \includegraphics[width=0.32\textwidth]{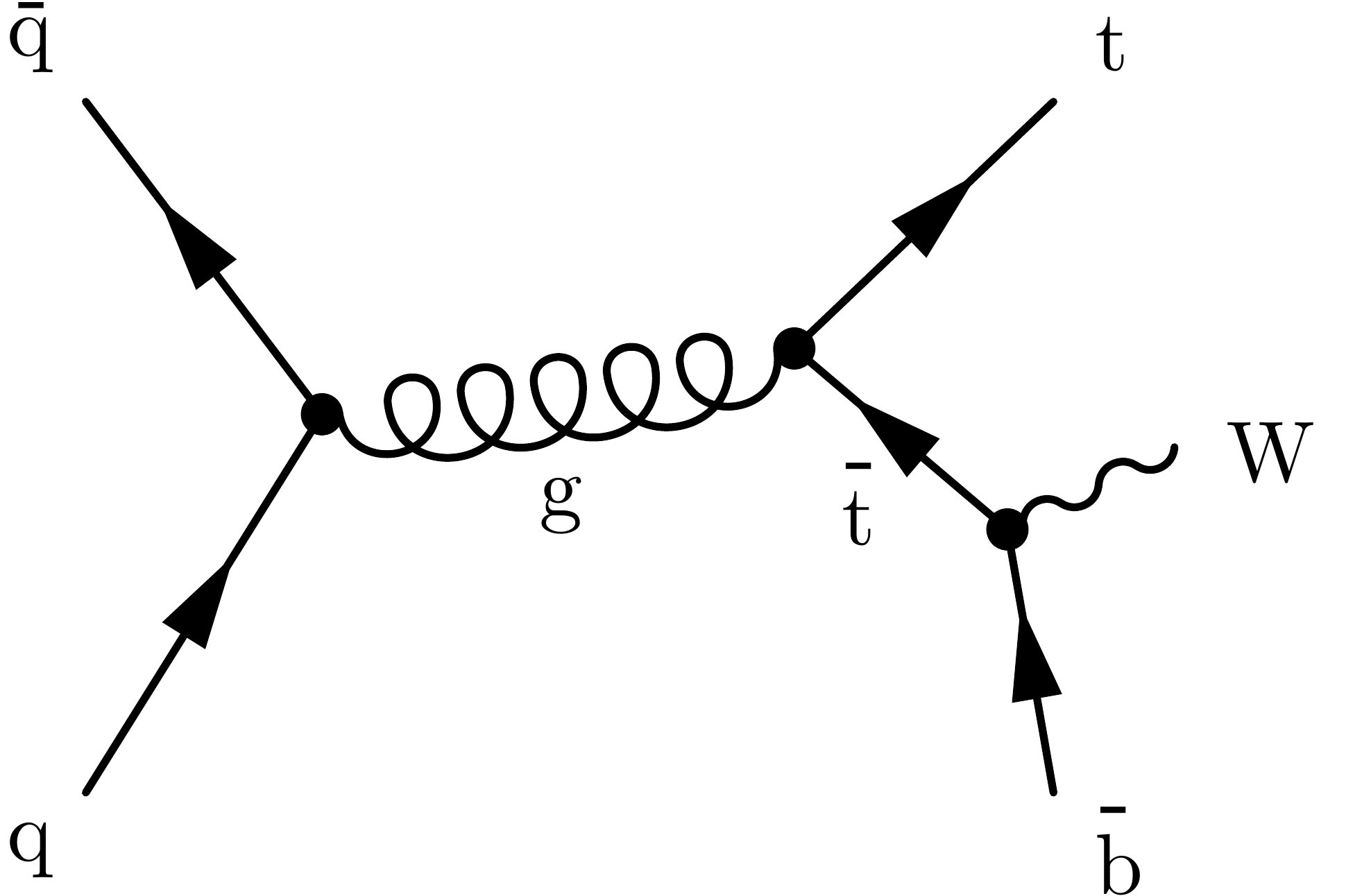}
    \caption{Feynman diagrams for \tW single top quark production at NLO that are removed from the signal definition in the DR scheme. The charge-conjugate modes are implicitly included.}
    \label{fig:diagsremoved}
\end{figure}

We rely on MC simulations to estimate the contributions from both signal and background processes. The \tW signal samples are simulated at NLO in QCD using \POWHEG v2~\cite{bib:powheg2,Frixione:2007vw,Nason:2004rx}. Two schemes are proposed to describe the \tW signal: ``diagram removal'' (DR)~\cite{Frixione:2008yi}, where all NLO diagrams that are doubly resonant (\ie that can have two top quarks on-shell), such as those in Fig.~\ref{fig:diagsremoved}, are excluded from the signal definition; and ``diagram subtraction'' (DS)~\cite{Frixione:2008yi,PhysRevD.61.034001}, in which the differential cross section is modified with a gauge-invariant subtraction term, which locally cancels the contribution of \ttbar diagrams. The DR scheme is used as the nominal model in this analysis. However, the difference in the results obtained for the two schemes is also evaluated. Both DR and DS schemes are used for comparison with the differential particle-level result. In addition, we consider signal samples generated with \MGvATNLO v2.6.5~\cite{Alwall:2014hca} to compare, at the particle level, the predictions that use the DR and DS schemes. Two other derivations from those approaches are also considered: the so-called ``DR2'' approach that includes the terms corresponding to the interference between \tW and \ttbar processes, and an alternative way of implementing DS (later referenced as ``DS dyn.''), where a dynamic factor is used to model the top quark resonance, instead of a fixed one~\cite{Demartin:2016axk}. The NLO QCD setup in \POWHEG v2 is also used to simulate \ttbar events. The \dy (DY) background samples are simulated at NLO in QCD using \MGvATNLO v2.2.2, except in 2017, where it is simulated at LO in QCD using \MGvATNLO v2.4.2. The contributions from $\PW\PW$, $\PW\PZ$, and $\PZ\PZ$ (referred to as \VV) processes are simulated at NLO in QCD with \POWHEG v2 or \MGvATNLO v2.2.2. Other contributions from \PW, \PZ, and \PGg boson production in association with \ttbar events (referred to as \ttV) are simulated at NLO in QCD using \MGvATNLO v2.2.2. The small cross section processes of triboson production with different \PW, \PZ, and \PGg combinations are grouped together with the \VV and \ttV contributions and simulated at NLO in QCD with \MGvATNLO v2.2.2.

The lepton+jets events in the \ttbar samples are used to estimate the contribution to the background from events with a jet incorrectly reconstructed as a lepton (electron or muon). Since the latter background contributions contain a lepton candidate that does not originate from a leptonic decay of a gauge boson, they are labelled \nonWZ.

The NNPDF 3.1 NNLO~\cite{Ball:2017pdf} PDF set is used in the simulation of all samples, except for the 2016 non-\ttbar backgrounds for which 3.0 NNPDF NLO~\cite{Ball:2014uwa} is applied. The generators are interfaced in all cases with \PYTHIA v8.230~\cite{Sjostrand:2014zea}, which is used to model the hadronisation and parton showering (PS). The underlying event is modelled with the CP5 tune~\cite{Sirunyan:2019dfx} in all of the 2017--2018 samples. For 2016, the CP5 tune is used for the signal and \ttbar background samples, whereas for the rest of the samples the CUETP8M1 tune~\cite{Khachatryan:2015pea} is taken. For comparison at the particle level, another signal sample where the \POWHEG v2 generator is interfaced with \HERWIGpp v2.7.1 (2016) and \HERWIG 7 (2017--2018)~\cite{Bahr:2008pv,Bellm:2015jjp,Bellm:2017bvx} is used. This uses the EE5C tune~\cite{Gieseke:2012ft} for 2016 and CH3~\cite{CMS:2020dqt} for 2017--2018. For the samples generated with \MGvATNLO at LO (NLO) accuracy, double counting of partons from the matrix element calculations and PS described by \PYTHIA v8.230 is removed using the \textsc{mlm}~\cite{Alwall:2007fs} (\textsc{FxFx}~\cite{Frederix:2012ps}) matching scheme. The nominal \mtop is set to 172.5\GeV for all samples. For both the \tW signal and \ttbar background alternative samples were generated to estimate systematic uncertainties, which are obtained from the same generator (\POWHEG v2) and PS simulation (\PYTHIA v8.230). These uncertainties are described in detail in Section~\ref{sec:uncs}.

The \GEANTfour package~\cite{Agostinelli:2002hh} is used to simulate the CMS detector response for all simulated samples. To compare with the measured data, the event yields in the simulated samples are normalised to the integrated luminosity using their theoretical cross sections. These are taken from calculations at NNLO for DY~\cite{PhysRevD.86.094034}, approximate NNLO for \tW events~\cite{Kidonakis:2015nna}, and NLO for diboson events~\cite{bib:mcfm:diboson}. For the simulated \ttbar sample, the full NNLO plus next-to-next-to-leading-logarithmic calculation~\cite{Czakon:2013goa}, performed with the \textsc{top++} 2.0 program~\cite{Czakon:2011xx}, is used. The PDF uncertainty is added in quadrature to the uncertainty associated with the strong coupling constant (\alpS) to obtain a \ttbar production cross section of $832\,^{+20}_{-29}\,\text{(scale)}\pm 35\,\text{(PDF+\alpS)}\unit{pb}$, for $\mtop = 172.5\GeV$.

The simulated samples include additional \pp interactions in the same or nearby bunch crossings (pileup). A reweighting is applied in simulations to match the distribution of bunch crossings observed in the data. Assuming a total inelastic \pp cross section of 69.2\unit{mb}~\cite{CMS:2018mlc}, the average number of pileup interactions per bunch crossing are 23, 33, and 32 in 2016, 2017, and 2018, respectively.

\section{Event selection}
\label{sec:eventselection}
In the SM, top quarks decay nearly 100\% of the time into a \PW boson and a bottom quark. The analysis described here uses events in the \emu final state, in which the \PW boson from the decay of the top quark and the \PW boson produced in association with the top quark both decay leptonically, one into an electron and a neutrino, and the other into a muon and a neutrino. This leads to a final state composed of two different-flavour leptons with opposite electric charge, one jet resulting from the fragmentation of a bottom quark, and two neutrinos.

Events are required to pass either a dilepton or single-lepton trigger with isolation requirements~\cite{CMS:2016ngn}. The dilepton triggers require events to contain either one electron with transverse momentum $\pt>12\GeV$ and one muon with $\pt>23\GeV$, or one muon with $\pt>8\GeV$ and one electron with $\pt>23\GeV$. The single-lepton triggers with one electron (muon) with $\pt>27$ (24), 35 (24), and 32 (24)\GeV for 2016, 2017, and 2018 are used to increase the efficiency. The combined trigger efficiency is measured using data events which pass the selection criteria given below, and which were collected with triggers based on the \pt imbalance in the event. It is found to be $\approx$98\%. The trigger efficiency in simulated events is corrected to match that observed in data.

The particle-flow (PF) algorithm~\cite{CMS:2017yfk} aims to reconstruct and identify each individual particle in an event with an optimised combination of information from the various elements of the CMS detector. The primary vertex, which is the vertex corresponding to the hardest scattering in the event, is evaluated using tracking information alone, as described in Section 9.4.1 of Ref.~\cite{CMS-TDR-15-02}. Further requirements are imposed on the reconstructed and identified lepton and jet candidates obtained from the PF algorithm. Electrons~\cite{Khachatryan:2015hwa} and muons~\cite{Sirunyan:2018fpa} in the event are required to be well isolated and have $\pt > 20\GeV$ and $\abseta < 2.4$. Additional criteria are imposed on the impact parameter of the leptons in order to ensure that they come from the primary vertex. Electron candidates in the transition region between the barrel and endcap calorimeters, corresponding to $1.444 < \abseta < 1.566$, are ignored because the electron reconstruction in this region is not optimal.

Events with \PW bosons decaying into \PGt leptons are considered as signal only if the \PGt leptons decay into electrons or muons that satisfy the selection requirements. In events with more than two leptons passing the selection, the two with the largest \pt are kept for further study. Jets are reconstructed from the PF candidates using the anti-\kt clustering algorithm~\cite{Cacciari:2008gp, Cacciari:2011ma} with a distance parameter of 0.4.
Jet energy corrections, derived from simulation, are applied so that the average response to jets matches the particle-level jets~\cite{CMS:2016lmd}. In situ measurements of the momentum balance in dijet, photon+jet, {\PZ}+jet, and multijet events are used to account for residual differences in the jet energy scale (JES) between data and simulation. In addition, jet energy resolution (JER) is corrected to reproduce that obtained from data~\cite{CMS:2019ctu}.

Jets are required to have $\pt > 30\GeV$ and $\abseta < 2.4$. Jets with \pt between 20 and 30\GeV and $\abseta < 2.4$ are referred to as ``loose jets''. The differences in these lower-\pt jets between the \tW and \ttbar distributions can be exploited for their separation. Jets are identified as coming from the fragmentation of bottom quarks (\PQb jets) using the DeepJet algorithm~\cite{CMS-DP-2018-058, Bols:2020bkb}, with an operating point that yields identification efficiencies of $\approx$70\% and misidentification probabilities of about 1\% for light-quark and gluon jets. The missing transverse momentum vector \ptvecmiss is defined as the negative vector momenta sum of all reconstructed PF candidates, which include the jet energy corrections described above in an event, projected onto the plane perpendicular to the direction of the beam axis. Its magnitude is referred to as \ptmiss.

\begin{figure}[htbp!]
    \centering
    \includegraphics[width=0.45\textwidth]{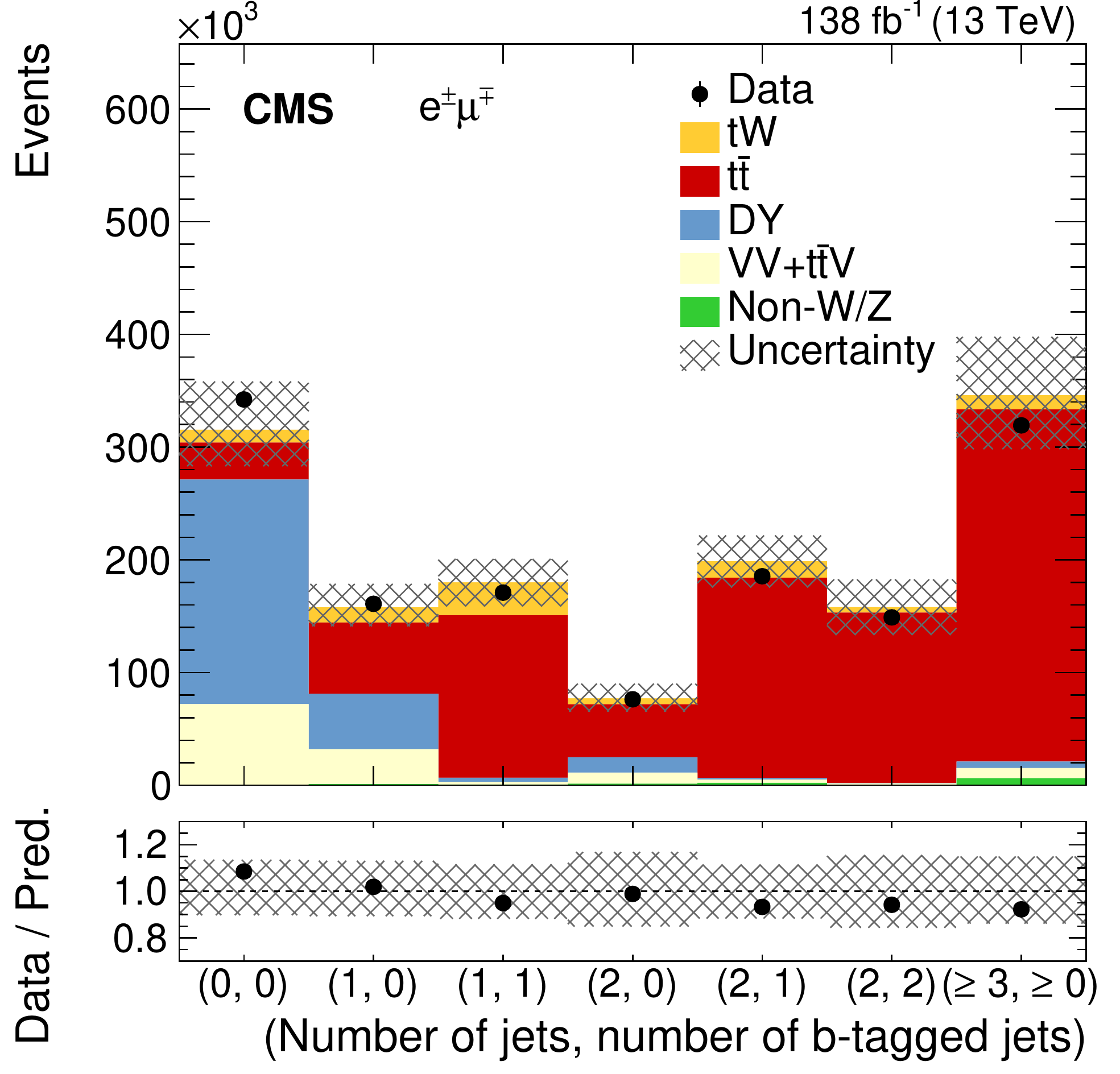}%
    \hspace*{0.05\textwidth}
    \includegraphics[width=0.45\textwidth]{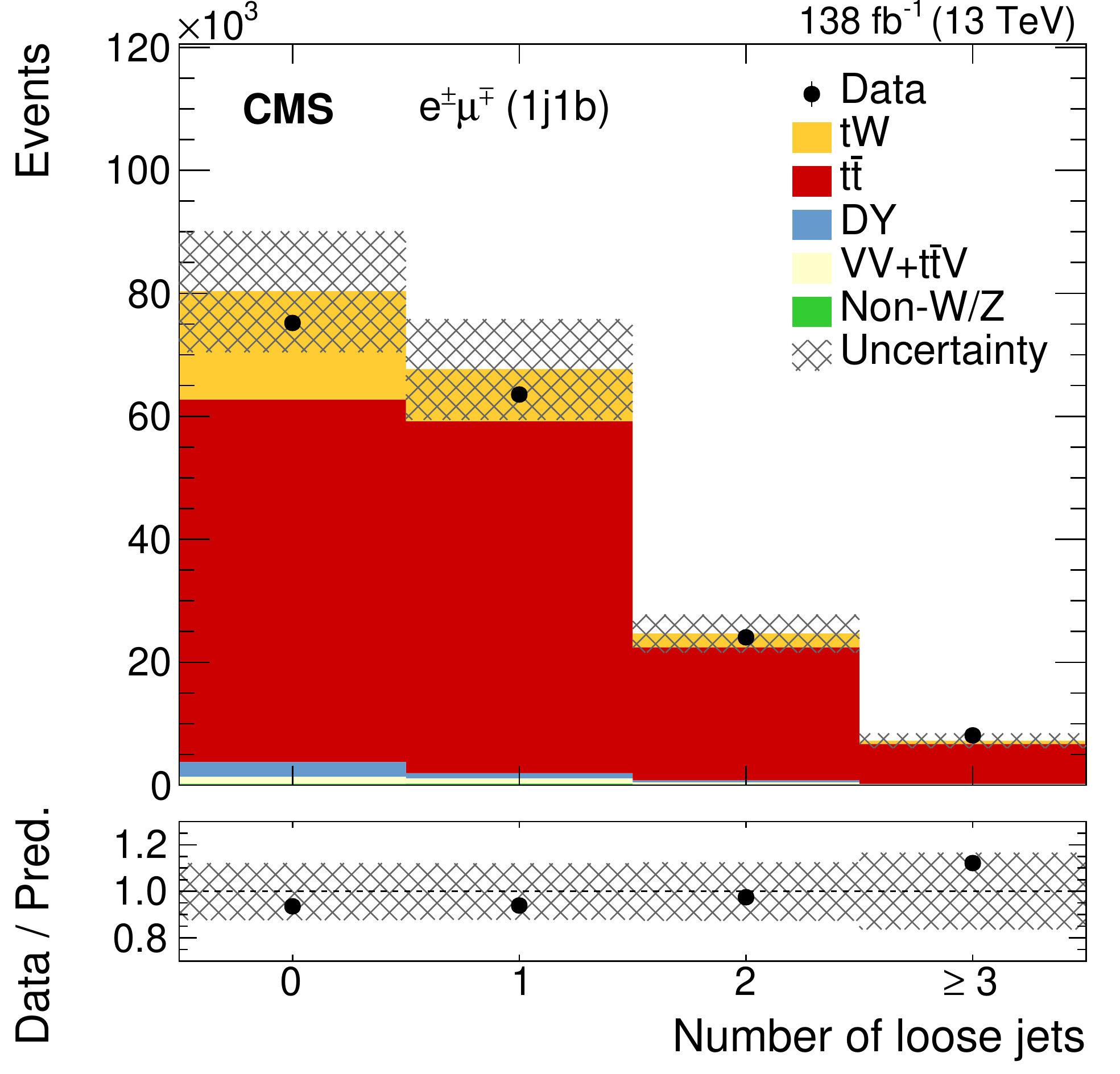}
    \caption{Left: the number of events from data (points) and predicted from simulation (coloured histograms) before the fit in the \emu sample as a function of the number of jets and \PQb-tagged jets. Right: the number of loose jets per event in the \emu sample from the \njnb{1}{1} region. The vertical bar on the points shows the statistical uncertainty in the data. The hatched band represents the sum of the statistical and systematic uncertainties before the fit. The lower panels show the ratio of data to the sum of the expected yields. The MC simulations are normalised to their theoretical cross section values as described in Section~\ref{sec:montecarlo}.}
    \label{fig:nloosejetsnjetnbtag}
\end{figure}

Events are selected as belonging to the \emu final state if the two leptons with highest \pt passing the above selection criteria  are an electron and a muon of opposite charge. The highest \pt (leading) lepton is required to have $\pt > 25\GeV$. To reduce the contamination from DY production of \PGt lepton pairs with low dilepton invariant mass, the minimum invariant mass of all pairs of identified leptons (including leptons beyond the leading two) is required to be greater than 20\GeV. The remaining events are classified by the number of jets and identified \PQb jets in the event, as shown in Fig.~\ref{fig:nloosejetsnjetnbtag} (left). In the following, the notation \njnb{n}{m} represents events with exactly $n$ jets where $m$ of them are identified as \PQb jets. The most signal-enriched region is the \njnb{1}{1}, but the contribution from \tW is still only 20\% that of \ttbar. For the inclusive measurement, we take advantage of the information from various regions (\njnb{1}{1}, \njnb{2}{1}, and \njnb{2}{2}), whereas for the differential measurement, only the \njnb{1}{1} region is used. An additional selection criterion is applied to enhance the signal-to-background ratio in the differential measurement. Figure~\ref{fig:nloosejetsnjetnbtag} (right) shows the distribution of the number of loose jets in the \njnb{1}{1} events. The signal-to-background ratio is larger for events with zero loose jets. To minimise the relative contribution from the \ttbar background, the events in the \njnb{1}{1} region with zero loose jets are used for the differential measurement. Signal events contribute up to 22\% of the total expected events in that region.

\section{Methodology for the inclusive measurement}
\label{sec:inclusive}
After the baseline event selection is performed, the presence of \ttbar events is considerably larger than the \tW signal in all event categories, as can be seen in Fig.~\ref{fig:nloosejetsnjetnbtag} (left). The region with the best signal-to-background ratio is \njnb{1}{1}, which consists mainly of \tW and \ttbar events. This region is used together with the \njnb{2}{1}, which also contains a significant \tW contribution, and \njnb{2}{2} in a maximum likelihood (ML) fit to extract the \tW signal. The \njnb{2}{2} region is dominated by \ttbar events and is used in the fit to constrain this background.

As there is no single observable that gives strong discrimination between \ttbar and \tW events, two independent boosted decision trees (BDTs~\cite{FREUND1997119,tmva}), one for the \njnb{1}{1} region and the other for the \njnb{2}{1} region, are trained to discriminate between the \tW signal and \ttbar background. The BDTs outperform single-tree classifiers~\cite{decision_trees} by training a set of trees (forest) and taking their weighted vote as the prediction. Each tree is derived from the same training ensemble by reweighting their events to mitigate the statistical fluctuations and increase the overall stability. In this analysis, the BDT implementation is provided by the \textsc{tmva} package~\cite{tmva}, using gradient boost~\cite{tmva} as the boosting algorithm. The BDTs are trained and tested using a set of simulated samples that are statistically independent from the ones used in the signal extraction.

The input variables used in the BDTs are chosen depending on how well the MC simulation models the data and on their discrimination power. For the BDT in the \njnb{1}{1}, the list of variables used in the training in order of importance are:
\begin{itemize}
    \item $\pt (\Pepm, \PGmmp, \PQj)$: the magnitude of the transverse momentum of the dilepton + jet system.
    \item $C(\Pepm, \PGmmp, \PQj)$: centrality, which is defined as $\sin{\theta}$, where $\theta$ is the polar angle of the total momentum of the system.
    \item $m(\Pepm, \PGmmp, \PQj, \ptvecmiss)$: invariant mass of the dilepton + jet + \ptvecmiss system.
    \item Leading loose jet \pt: if there are no loose jets, this variable is set to 0.
    \item Jet \pt.
    \item Presence of loose jets in the event: the result is either yes or no.
\end{itemize}

\begin{figure}[htbp!]
    \centering
    \includegraphics[width=0.45\textwidth]{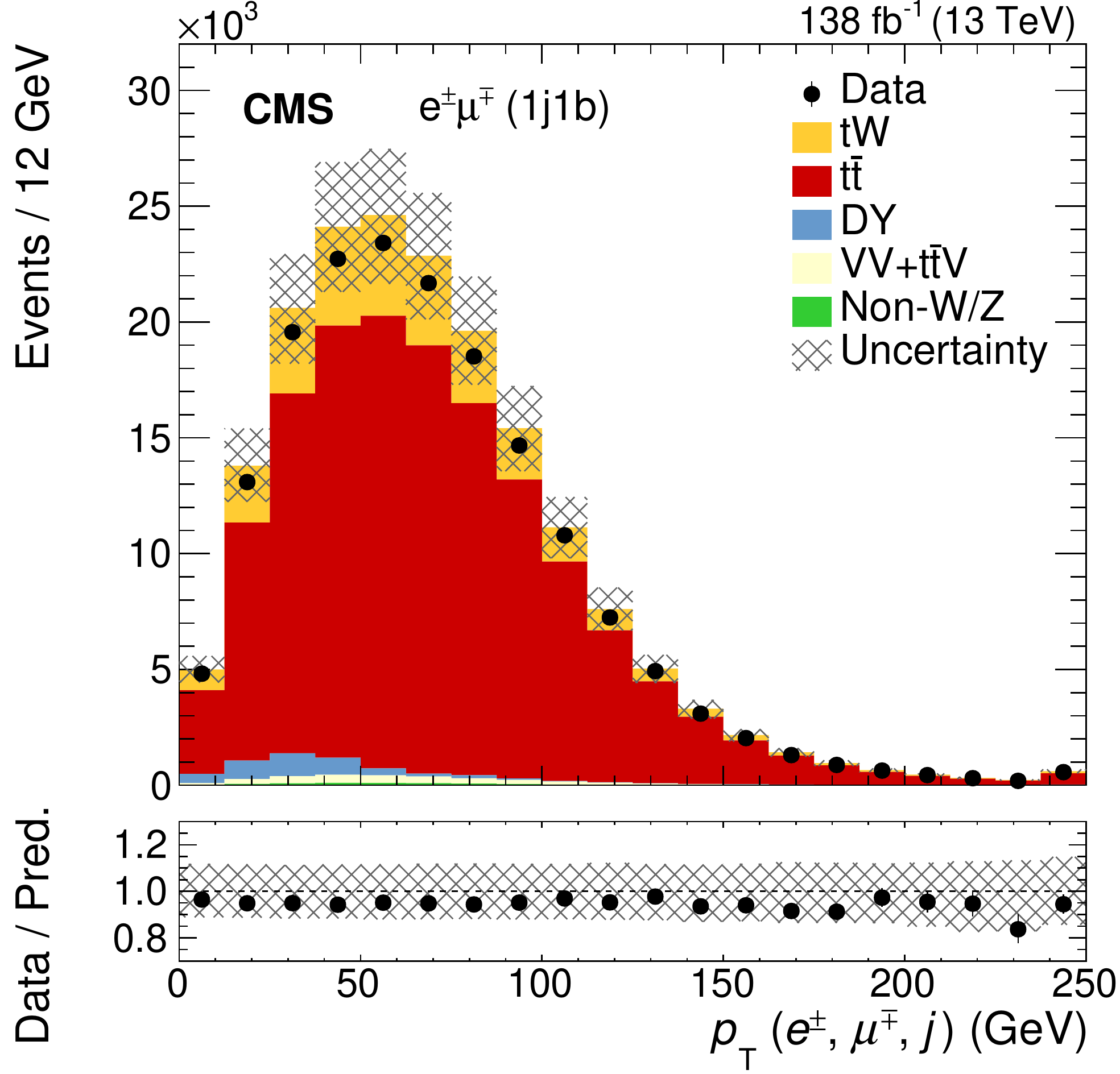}%
    \hspace*{0.05\textwidth}%
    \includegraphics[width=0.45\textwidth]{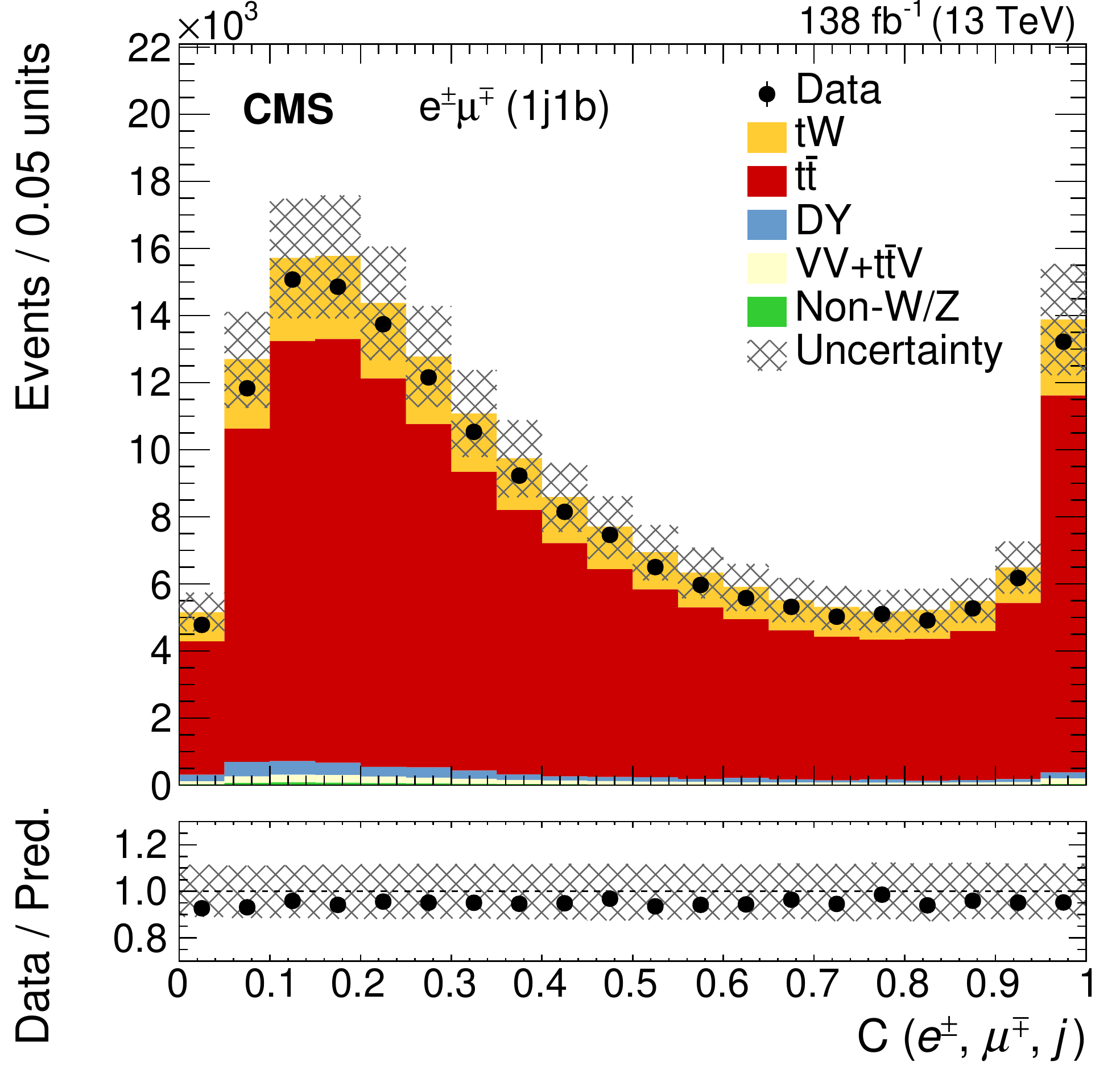}\\
    \includegraphics[width=0.45\textwidth]{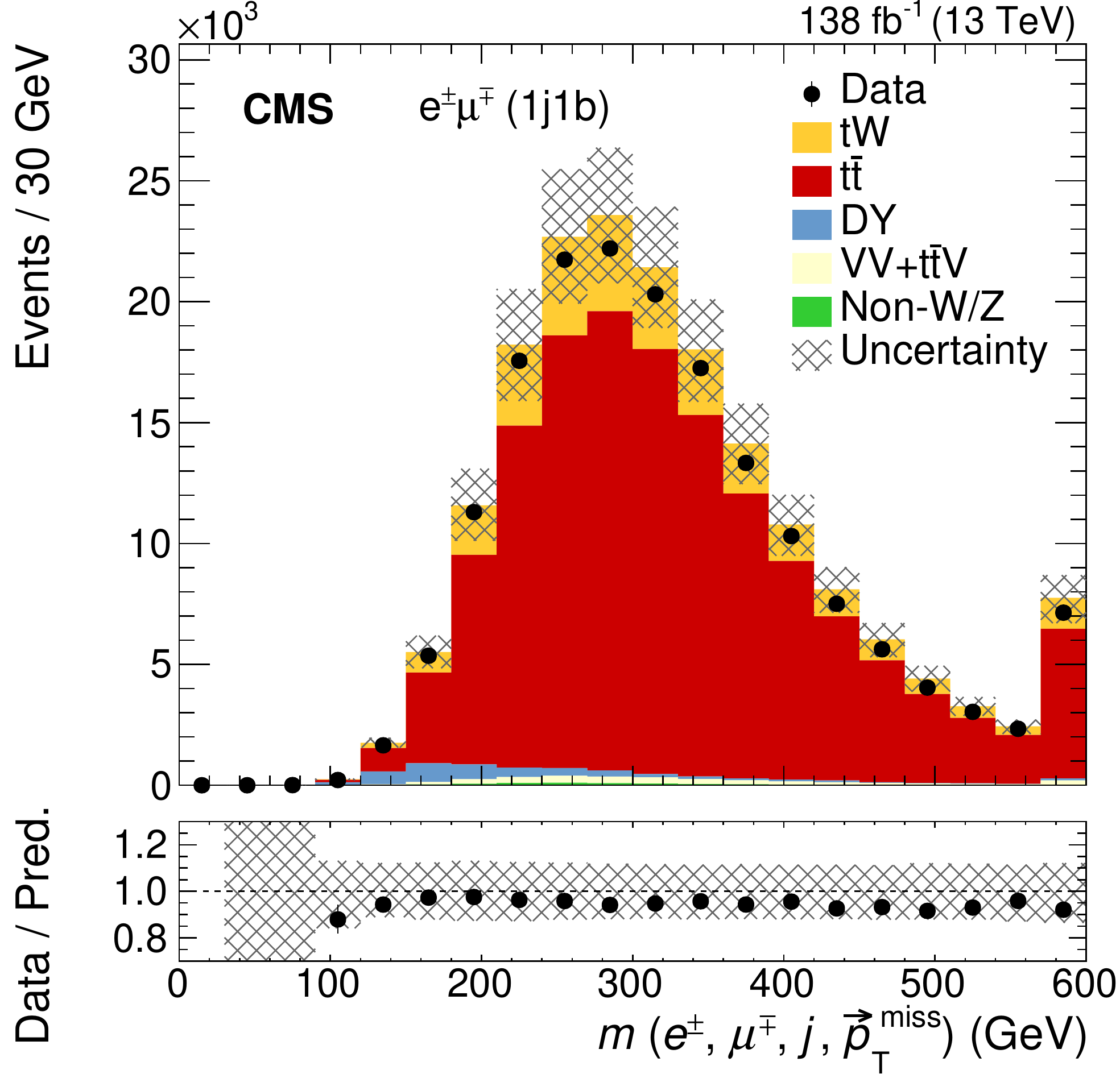}%
    \hspace*{0.05\textwidth}%
    \includegraphics[width=0.45\textwidth]{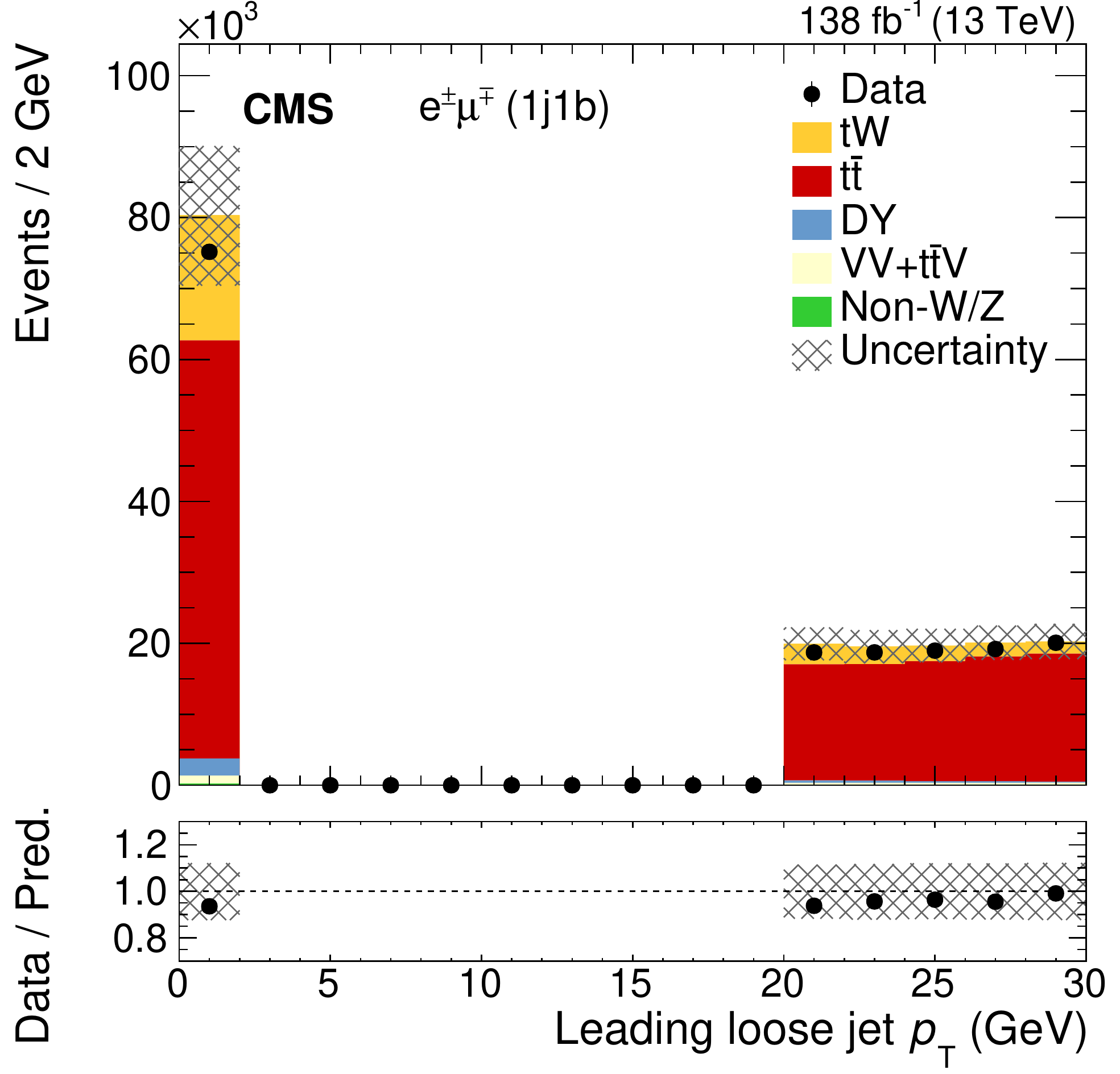}
    \caption{Distributions from data (points) and MC simulations (coloured histograms) of the four most discriminating variables used for the BDT training of the \njnb{1}{1} region: (upper left) the magnitude of the transverse momentum of the dilepton + jet system; (upper right) the centrality, defined in the text, of the dilepton + jet system; (lower left) the invariant mass of the dilepton + jet + \ptvecmiss system; and (lower right) the \pt of the leading loose jet. The last bin of each distribution includes the overflow events. The first bin in the lower right plot contains events with 0 loose jets.  The vertical bars on the points give the statistical uncertainty in the data, and the grey bands represent the sum of the statistical and systematic uncertainties in the MC predictions. The lower panels show the ratio of the data to the sum of the MC predictions. The MC simulations are normalised to their theoretical cross section values as described in Section~\ref{sec:montecarlo}.}
    \label{fig:InputVariables1j1bData-MC}
\end{figure}

The order of importance is determined by counting how often each variable is used to split the decision tree nodes. The counts are weighted by the square of the separation gain achieved by the variable and by the number of events in the node. Figure~\ref{fig:InputVariables1j1bData-MC} shows the data-MC agreement of the four most discriminating variables in the BDT for \njnb{1}{1}, where good agreement is observed. The same applies for the remaining distributions of the presence of loose jets and the jet \pt, and for the input variables of the BDT trained in the \njnb{2}{1} region.

The input variables listed in order of importance in the BDT for \njnb{2}{1} are:
\begin{itemize}
    \item $\DR(\Pell_1, \PQj_1)$: separation in \etaphi space between the leading lepton and leading jet, where $\varphi$ is the azimuthal angle.
    \item $\DR(\Pell_{12}, \PQj_{12})$: separation in \etaphi space between the dilepton and dijet systems.
    \item Second-highest jet \pt (subleading).
\end{itemize}
Three distributions are considered in the ML fit: the BDT output distributions in the \njnb{1}{1} and \njnb{2}{1} regions, and the \pt distribution of the subleading jet in the \njnb{2}{2} region. This last variable is sensitive to JES variations and is useful in constraining this systematic uncertainty. The binning of the BDT output distribution is chosen such that each bin contains about the same number of \ttbar events. This avoids the presence of low-statistic bins in the background estimation, helping to constrain the systematic uncertainties. The fit is performed simultaneously with the three regions. The uncertainties are included using nuisance parameters, one for each source of systematic uncertainty, correlated across all regions, which parameterise the effect of the given source on the expected signal and background yields.

The likelihood used in this ML fit, $\mathcal{L}(\vecn | \mu,\vectheta)$, is a function of the signal strength $\mu$, defined as the ratio of the measured and expected SM cross sections $\mu = \tWxsec/\tWxsecSM$, the observed number of events in each bin \vecn, and a set of nuisance parameters \smash{\vectheta} that parameterise the systematic uncertainties. It is constructed as the product of Poisson probabilities corresponding to the number of events in each bin of the distributions. Additionally, the systematic uncertainties are introduced in the likelihood multiplied by the prior $p_j(\thetaj)$ of each nuisance parameter \thetaj. In this analysis, a log-normal probability density function is used for nuisance parameters affecting the normalisation of different signal and background processes, and a Gaussian distribution is employed for the shape uncertainties. The best fit value for $\mu$ is obtained by maximising the likelihood function with respect to all its parameters. The ML fits are implemented with software based on \textsc{RooStats}~\cite{RooStats}. The MC statistical uncertainties are incorporated using the Barlow--Beeston method~\cite{Barlow:1993dm,Conway:2011in}. Other uncertainties that might affect both the normalisation and shape of the distributions are introduced using specific nuisance parameters~\cite{ATLAS:2011tau,Conway:2011in} with a Gaussian function prior.

\section{Methodology for the differential measurement}
\label{sec:differential}
The spectra of observables are distorted by the response and acceptance of the detector. Unfolding techniques must, therefore, be used to determine the actual distributions without the detector effects so that these can be directly compared with theoretical predictions. The data affected by these distortions are said to be at detector level. The parton level is defined by the particles produced after the generation of the hard-scattering process. If the information from the PS and hadronisation simulations is added, this then gives the particle level.

The measured distributions are unfolded from the detector level to the particle level. Unfolding to particle level instead of parton level reduces the migration and efficiency corrections, and allows the fiducial region to be defined in close correspondence with the event selection of the analysis.

The identification of particle-level objects is summarised in Table~\ref{tab:partLevel}. These objects are constructed using stable (\ie with a lifetime larger than 30\unit{ps}) generated particles following the conventions given in Ref.~\cite{pseudotop}. Muons and electrons not coming from hadronic decays (prompt leptons) are ``dressed'' by taking into account the momenta of nearby photons within a $\DR < 0.1$ cone, where $\DR$ is the separation in \etaphi space between the muon or electron and the photon. Jets are clustered from all of the stable particles excluding prompt electrons, prompt muons, prompt photons, and neutrinos, using the anti-\kt algorithm with a distance parameter of $R=0.4$. The information of the intermediate hadrons and \PGt leptons is preserved inside the jets and used to determine whether a jet originates from the fragmentation of a heavy-flavour quark (bottom or charm) or whether it is a decay product of a \PGt lepton. With these requirements, a fiducial region is defined as described in Table~\ref{tab:fidspace}. Requiring exactly one \PQb jet reduces the potential to have events from the doubly-resonant diagrams. Those events, more affected by the interference between processes, are expected to have a larger jet multiplicity. The differences between the various models used to treat the interference are expected to be higher when the presence of events from doubly-resonant diagrams is larger, and vice versa. For the unfolded distributions in the fiducial region, as shown in Fig.~\ref{fig:particlefidbin1}, all MC simulations show very similar distributions. Therefore, this choice of fiducial region reduces these effects and the accompanying modelling uncertainty associated with the interference treatment (see Section~\ref{sec:uncs}).

\begin{table}[htbp!]
    \centering
    \topcaption{Selection requirements for particle-level objects.}
    \begin{tabular}{lcc}
        Object      & \pt ({\GeVns}) & \abseta \\
        \hline
        Muons       & $>$20 & $<$2.4 \\
        Electrons   & $>$20 & $<$2.4, excluding [1.444--1.566] \\
        Jets        & $>$30 & $<$2.4 \\
        Loose jets  & $>$20, $<$30 & $<$2.4 \\
    \end{tabular}
    \label{tab:partLevel}
\end{table}

\begin{table}[htbp!]
    \centering
    \topcaption{Definition of the fiducial region.}
    \begin{tabular}{lc}
        Number of leptons     & $\ge$2 \\
        Leading lepton \pt    & $>$25\GeV \\
        Invariant mass of all dilepton pairs  & $>$20\GeV\\
        Number of jets        & 1     \\
        Number of loose jets  & 0     \\
        Number of \PQb jets   & 1     \\
    \end{tabular}
    \label{tab:fidspace}
\end{table}

The signal extraction and the unfolding for the fiducial differential cross section measurement are performed with an ML fit designed as follows. The parameters of interest are the strengths of the signal process in each bin of the particle-level distribution. The signal sample is divided into as many contributions as there are particle-level bins. There is a 7\% contribution to the signal region from nonfiducial events. We treat these events as a background so the strength associated with them is not a parameter of interest. One nuisance parameter for each systematic uncertainty source is added to the fit.

\begin{figure}[htbp!]
    \centering
    \includegraphics[width=0.45\textwidth]{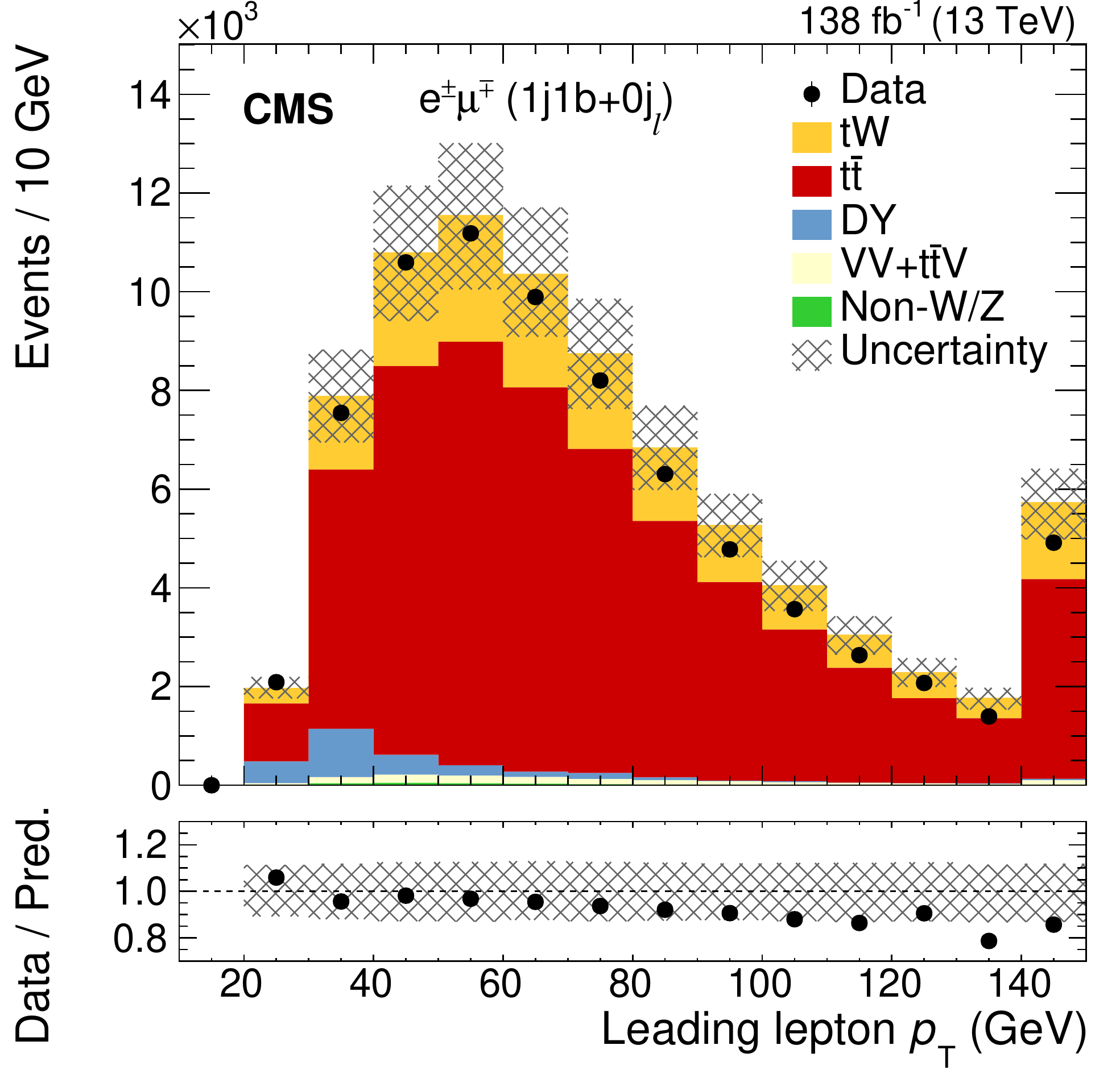}%
    \hspace*{0.05\textwidth}%
    \includegraphics[width=0.45\textwidth]{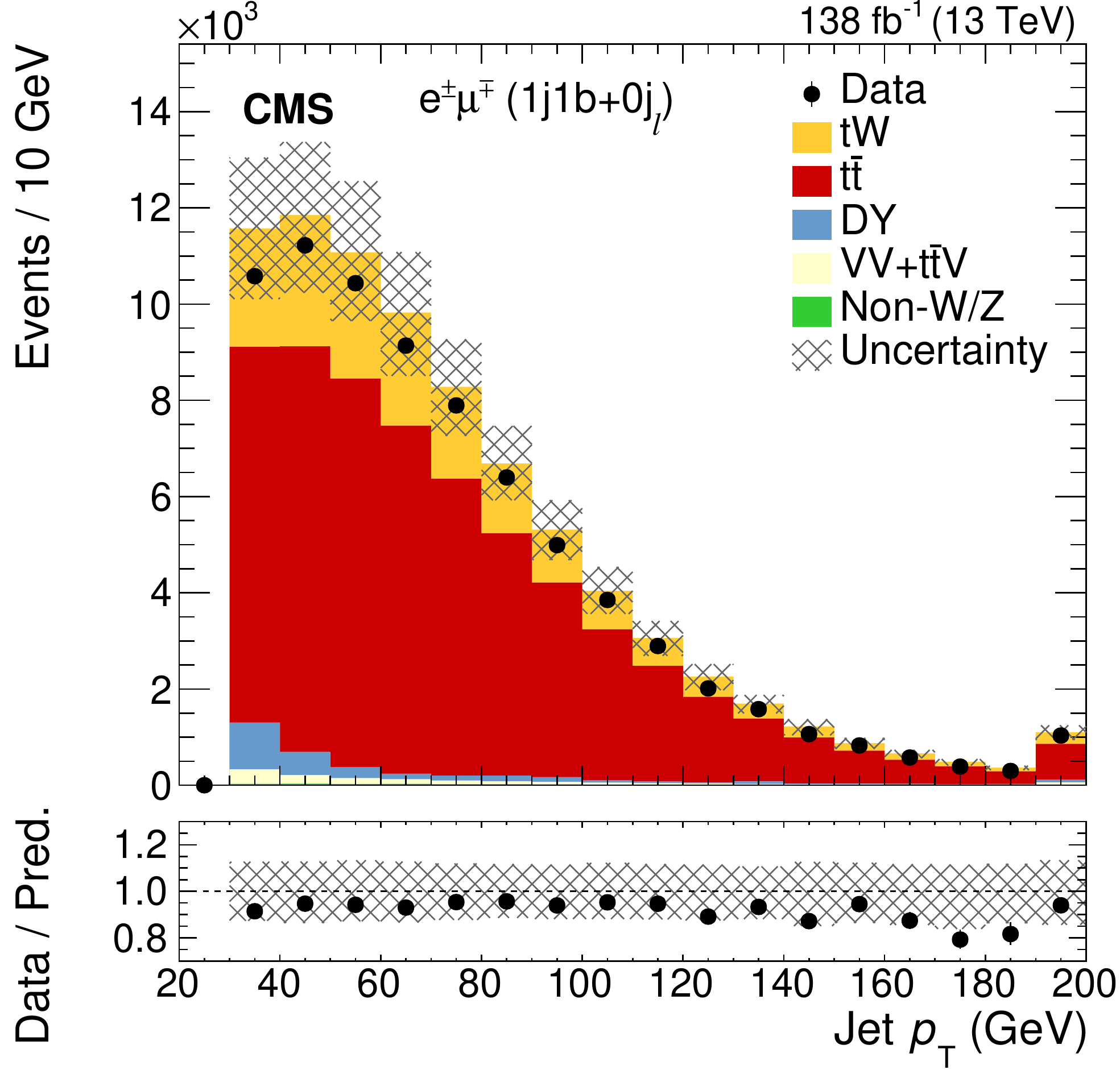} \\
    \includegraphics[width=0.45\textwidth]{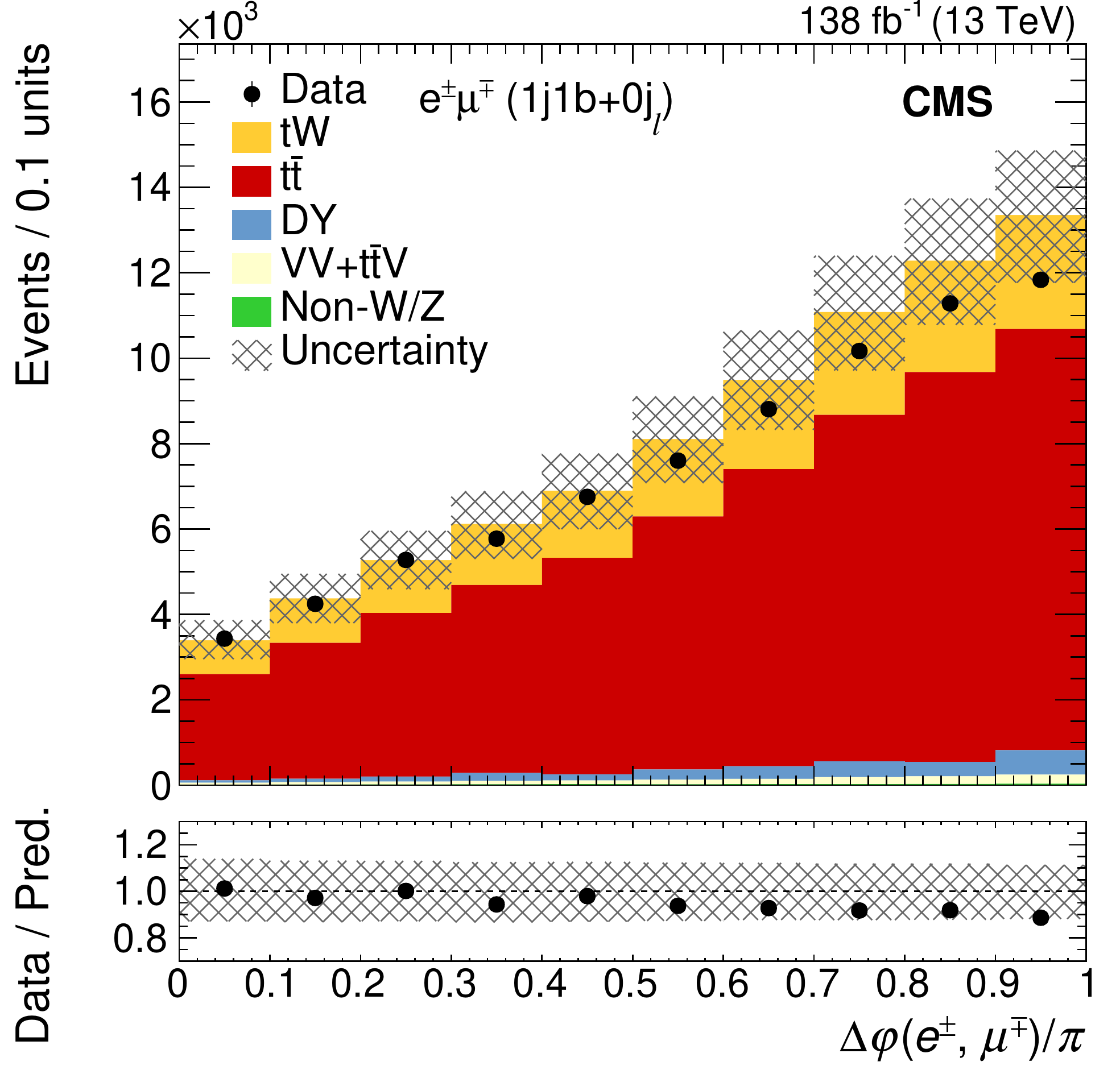}%
    \hspace*{0.05\textwidth}%
    \includegraphics[width=0.45\textwidth]{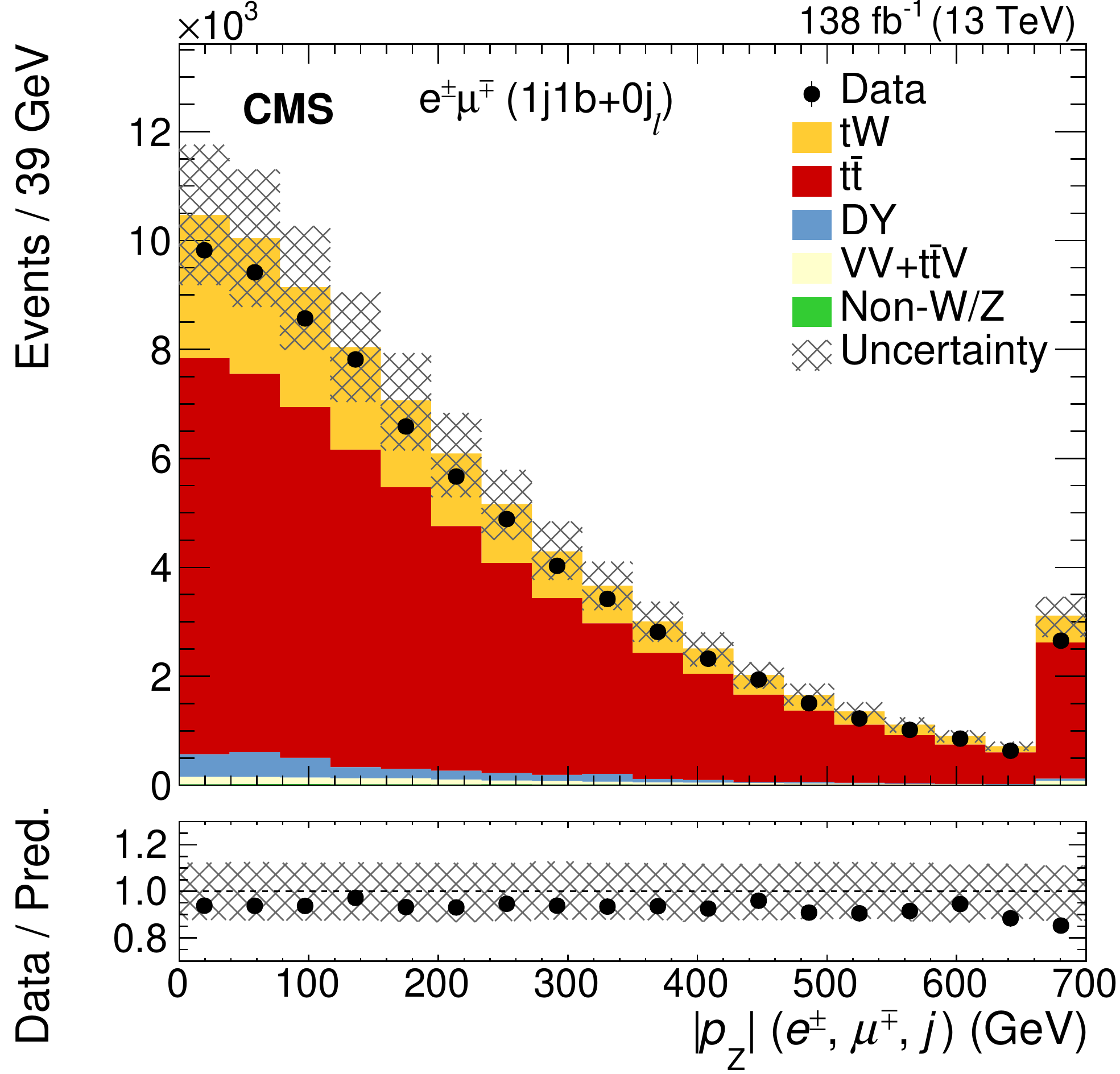} \\
    \includegraphics[width=0.45\textwidth]{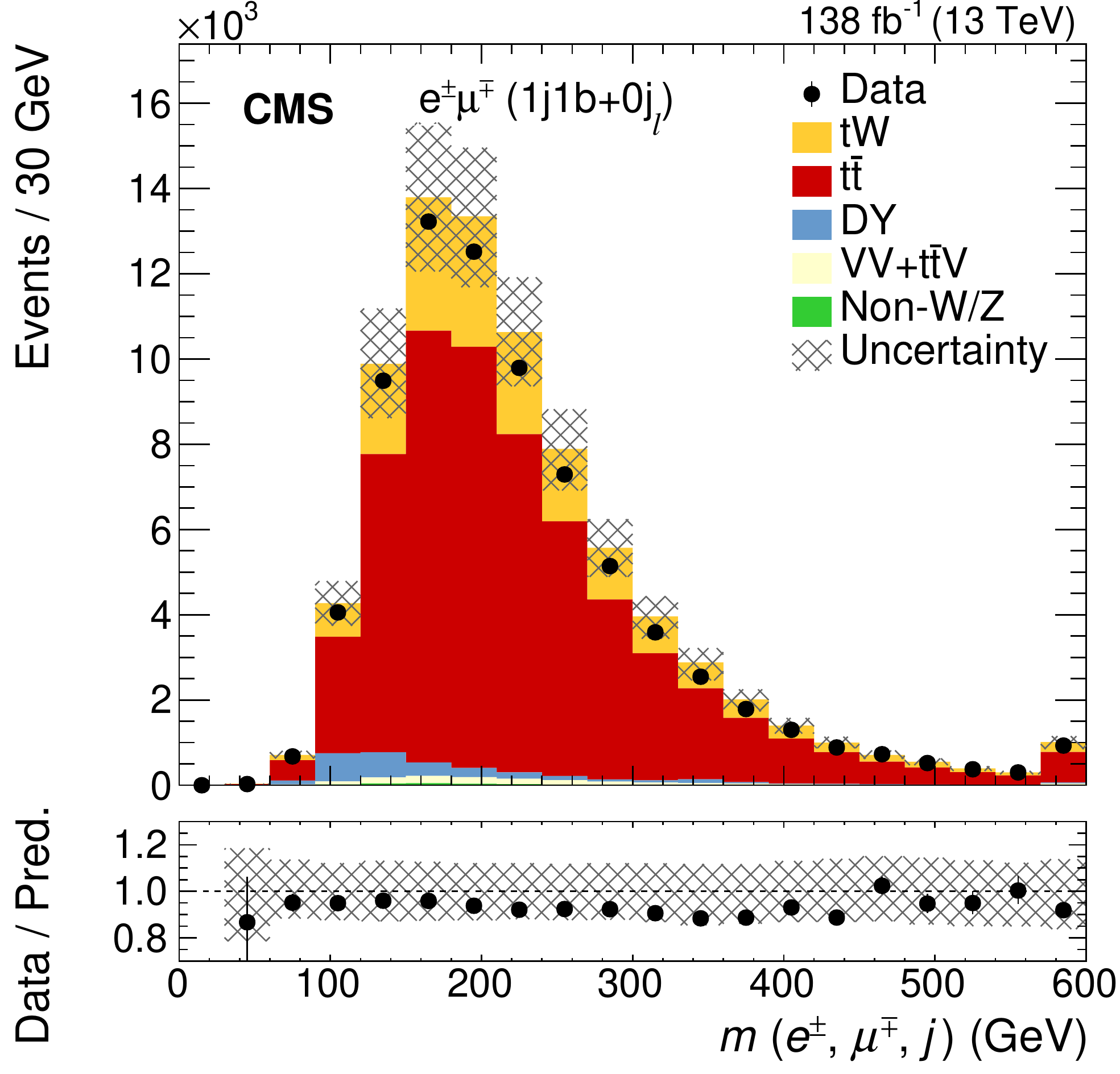}%
    \hspace*{0.05\textwidth}%
    \includegraphics[width=0.45\textwidth]{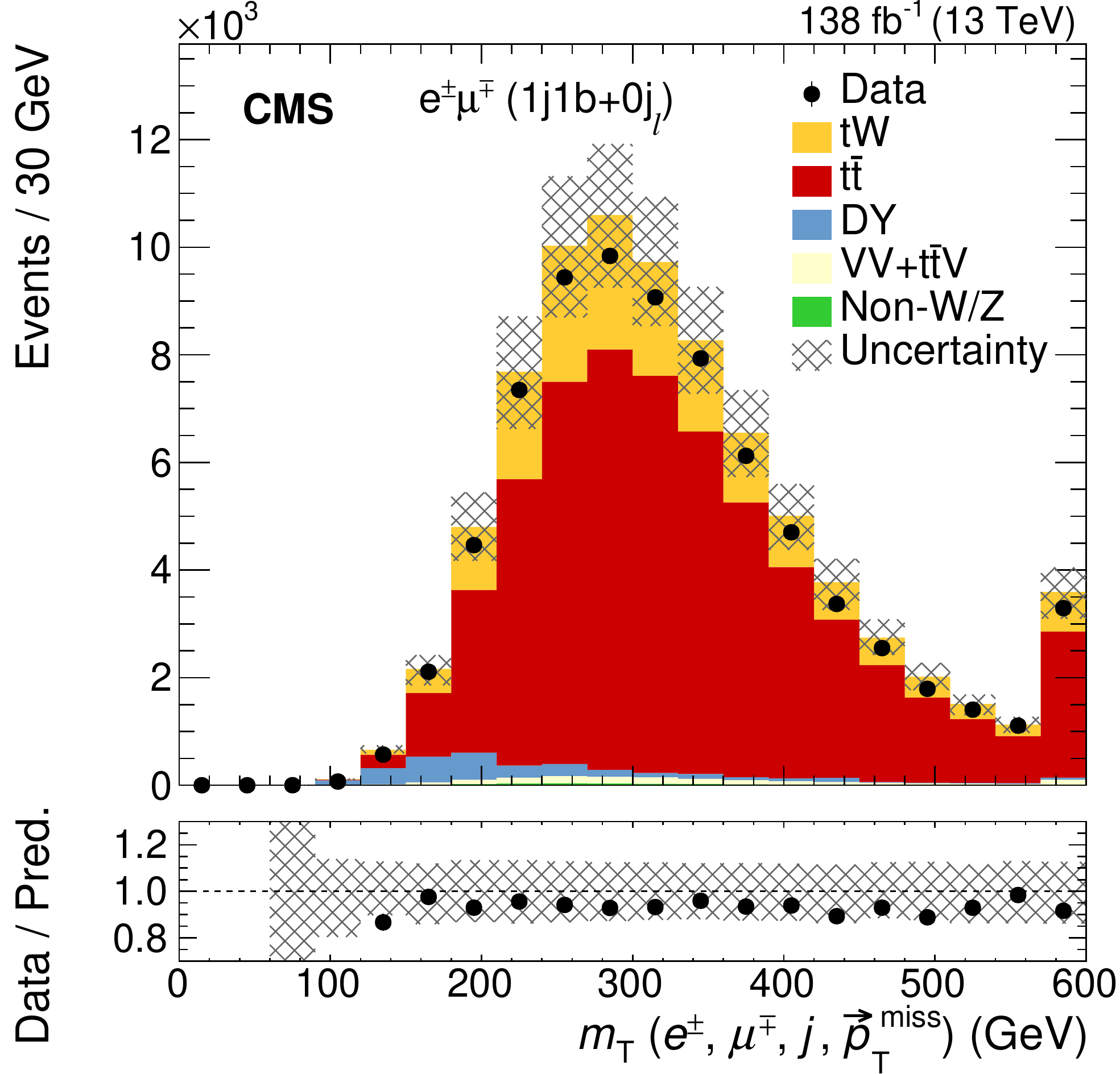}
    \caption{The measured distributions from data (points) and MC simulations (coloured histograms) of the six observables used to measure the \tW differential cross sections. Signal events in the \njnb{1}{1} region with 0 loose jets ($0\PQj_{l}$) are selected. The last bin of each distribution contains the overflow events. The vertical bars on the data show the statistical uncertainty. The hatched band displays the sum of the statistical and systematic uncertainties in the MC predictions before the fit. The lower panels show the ratio of the data to the sum of the MC expectations. The MC simulations are normalised to their theoretical cross section values as described in Section~\ref{sec:montecarlo}.}
    \label{fig:datamcdifferential}
\end{figure}

The differential cross section is measured as a function of the following physical observables:
\begin{itemize}
    \item leading lepton \pt;
    \item jet \pt;
    \item \deltaPhiVar: the azimuthal angle difference between the two leptons;
    \item \pzvar: the longitudinal momentum component of the dilepton + jet system;
    \item \invmassvar: the invariant mass of the dilepton + jet system; and,
    \item \transmassvar: the transverse mass of the dilepton + jet + \ptvecmiss system. For a collection of particles with transverse momentum \vecpTi, \mT is defined as:
\end{itemize}
\begin{equation}
    \mT = \sqrt{ \bigg( \sum_i \abs{\vecpTi} \bigg)^2 - \bigg| \sum_i \vecpTi \bigg|^2 }.
\end{equation}
The first two variables shown above provide information on the kinematic properties of the events. The \deltaPhiVar variable probes the kinematic and polarisation correlations between the top quark and \PW boson. The \pz distribution can be used to study the boost of the \tW system. The last two variables, the dilepton + jet invariant mass and \mT, are sensitive to the invariant mass of the \tW system. The distributions from the data and simulation for these six variables in the signal region are shown in Fig.~\ref{fig:datamcdifferential}. As in the case of Figs.~\ref{fig:nloosejetsnjetnbtag} and~\ref{fig:InputVariables1j1bData-MC}, overall there is good agreement within the uncertainties between the data and simulation, though the data are consistently lower than the predicted values.

The measurement is performed using all three years of data taking. The detector response is estimated using the response matrices, which are similar for the three years, with the matrices being almost diagonal. Thus, the measurement is performed directly using the combined data set and without need of regularisation. After the ML fit, the result is normalised to the fiducial cross section (obtained from the summation of the contents of the bins), and the bin width. The uncertainties are propagated taking into account the correlations across bins after the fit. The Asimov data set has been used to verify the closure and performance of the unfolding procedure.

\section{Systematic uncertainties}
\label{sec:uncs}
In addition to statistical uncertainties, the measurements of the inclusive and differential \tW cross sections are affected by systematic uncertainties that originate from both detector effects and theoretical assumptions. Each source of systematic uncertainty is assessed individually by suitable variations of the simulations or by variations of parameter values in the analysis within their estimated uncertainties. The systematic uncertainties are introduced in the ML fits as nuisance parameters. All experimental sources are applied to all processes.

The following text describes the sources considered for both the inclusive and differential analyses. Since we are considering data and simulation samples from different years, we indicate whether each uncertainty is correlated or not from year to year. In the case of the modelling uncertainties, we also indicate if they are correlated among the processes or not. Both normalisation and shape effects of all sources are considered apart from the background normalisation and integrated luminosity, which have only rate uncertainties. Effects of all sources in the estimation of the number of events that are not in the fiducial region are taken into account in the maximum likelihood fit, as well as their correlation with other uncertainties.

\subsection{Experimental uncertainties}
\begin{description}
    \item[JES and JER:] The uncertainty due to the limited knowledge of the JES and JER is determined by varying the scale and resolution within the uncertainties in bins of the jet \pt and $\eta$, typically by a few percent~\cite{CMS:2016lmd}. The JES and JER uncertainties are propagated to \ptvecmiss. The JES uncertainty sources are separated into various components that are correlated or uncorrelated across years in different groups. The JER uncertainty is uncorrelated across years.
    \item[Unclustered energy:] The effect of unclustered energy from the calorimeters on \ptvecmiss is taken into account through the momentum resolution of the various PF candidates~\cite{CMS:2017yfk,CMS:2015myp,CMS:2014pgm}. This is correlated across years.
    \item[\PQb tagging:] The uncertainties resulting from the \PQb tagging data-to-simulation scale factors (SFs) are assessed by varying them, within their uncertainties, for the \PQb, \PQc, light-flavoured, and gluon jets. These uncertainties vary with the \pt and $\eta$ of the jet and amount to approximately 2\% for \PQb-tagged jets and 10\% for misidentified jets~\cite{Sirunyan:2017ezt}, as determined in simulated \ttbar events. They are split into one correlated source (the systematic effect) across years, and one uncorrelated source (the statistical effect) per year for heavy-flavour (\PQb and \PQc) tagging and light-flavoured/gluon jet misidentification cases.
    \item[Trigger and lepton identification:] The uncertainties in the trigger and lepton identification and isolation are estimated by varying the data-to-simulation SFs by their uncertainties. These are about 0.7\% (trigger) and 1.5\% (lepton identification and isolation), with some dependence on the lepton \pt and $\eta$. The trigger uncertainties contain statistical sources (from both the MC simulations and data used) and systematical ones (estimating the effect of the event topology used to derive them and the consequences of selecting triggers based on the \pt imbalance of the events). The lepton identification and isolation scale factors are derived through the tag-and-probe method~\cite{CMS:2010svw} as a function of the lepton \pt and $\eta$. For muons, an additional uncertainty of 0.5\% is added in quadrature to account for the extrapolation from the phase space in which the isolation SFs are measured and the phase space for the analysis. The trigger and the statistical component of the muon identification uncertainties are uncorrelated across years, while the rest of the uncertainties are correlated. The identification and isolation uncertainties of muons are considered separately. In the case of electron identification and reconstruction uncertainties, they are taken as correlated across years.
    \item[Muon momentum scale:] To account for the uncertainties in the muon momentum scale, the momentum of the muons is varied by its uncertainties~\cite{Sirunyan:2018fpa}. These are uncorrelated across years. The experimental uncertainties in the electron energy (both scale and resolution) have been found to have a negligible effect and are thus not considered.
    \item[Pileup:] The uncertainty assigned to the number of pileup interactions in simulation is obtained by varying the inelastic \pp cross section within its uncertainty of 4.6\%~\cite{Aaboud:2016mmw}. This uncertainty is correlated across the three years.
    \item[Luminosity:] The uncertainty in the integrated luminosity is estimated to be 1.2, 2.3, and 2.5\% for 2016~\cite{CMS-LUM-17-003}, 2017~\cite{CMS-PAS-LUM-17-004}, and 2018~\cite{CMS-PAS-LUM-18-002}, respectively, with a combined uncertainty of 1.6\%. The uncertainty is partially correlated across the years.
    \item[L1 prefiring:] In 2016--2017, the L1 trigger from the electromagnetic calorimeter from the forward endcap region ($\abseta>2.4$) of the CMS detector showed a gradual shift in the timing of its inputs. This caused a effect known as ``prefiring'', where particles were assigned to previous collisions. This inefficiency only affects events having jets with high pseudorapidity ($2.4<\abseta<3.0$) and high \pt ($>$100\GeV). Events from these years are corrected through a reweighting determined from an unbiased data sample. An uncertainty equal to 20\% of the correction is taken. This uncertainty source is uncorrelated across the two years.
\end{description}

\subsection{Modelling uncertainties}
The impact of theoretical assumptions in the modelling is determined by repeating the analysis and replacing the standard \PHP\ \ttbar or \tW simulation by dedicated simulation samples with altered parameters, or by reweighting the nominal samples.
\begin{description}
    \item[Matrix element (ME) scales:] The uncertainty in the modelling of the hard process is only considered for \ttbar and \tW events and is assessed by changing independently the \muR and \muF scales in the \POWHEG sample by factors of 2 and 0.5 relative to their common nominal value. Unphysical variations of \muR and \muF, where the nominal values are shifted in opposite directions, are not considered. This uncertainty is correlated across years, and is performed separately for \ttbar and \tW events.
    \item[Parton shower (PS):] To take into account the PS uncertainties, different effects are studied:
    \begin{itemize}
        \item Underlying event: \PYTHIA parameters are tuned to the measurements of the underlying event~\cite{Sirunyan:2019dfx,Skands:2014pea}. These account for nonperturbative QCD effects. They are varied up and down in simulated \ttbar and \tW events. This variation is correlated across years and between \ttbar and \tW events.
        \item ME/PS matching: the uncertainty in the combination of the ME calculation with the PS in simulated \ttbar events is estimated from the variation, within its uncertainties, of the \POWHEG parameter $h_{\text{damp}} = \left(1.379 \, ^{+0.926}_{-0.505}\right) \mtop$~\cite{CMS-PAS-TOP-16-021,Sirunyan:2019dfx}. This parameter regulates the damping of real emissions in the NLO calculation when matching to the PS~\cite{Skands:2014pea}. This variation is correlated across years and is only considered for \ttbar events.
        \item Initial- and final-state radiation scales: the PS scale used for the simulation of the initial- and final-state radiations is varied up and down by a factor of two and only considered for \ttbar and \tW events. These variations are motivated by the uncertainties in the PS tuning~\cite{Skands:2014pea}. This variation is correlated across years for the final-state radiation in the \ttbar and \tW events, and is treated separately for initial-state radiation.
        \item Colour reconnection: the parameterisation of colour reconnection has been studied in Ref.~\cite{Argyropoulos:2014zoa}. A simulation including colour reconnection of early resonant decays (ERD) is used as the reference model. The uncertainties that arise from ambiguities in modelling are estimated by comparing with two alternative models of colour reconnection: a model with string formation beyond leading colour, and a model in which the gluons can be moved to another string~\cite{Colaboration:2022zut}. All models are tuned to measurements of the underlying event~\cite{CMS-PAS-TOP-16-021,Skands:2014pea}. The different models are included in the ML fits. This variation is correlated across years and between \ttbar and \tW events.
    \end{itemize}
    \item[PDFs and \alpS:] The uncertainty from the choice of PDF set is determined by reweighting the samples of simulated \ttbar and \tW events using 100 NNPDF3.1 replicas~\cite{Ball:2017pdf}. Since they represent the contents of a diagonalised Hessian matrix, the variations are summed quadratically. We then quadratically add the uncertainty in the \alpS parameter~\cite{Ball:2017pdf}. This uncertainty is correlated across years and between \ttbar and \tW events.
    \item [Top quark mass:] The nominal \mtop of 172.5\GeV is modified by $\pm$1\GeV and this is propagated to the MC simulations. This corresponds to twice the uncertainty in \mtop from CMS~\cite{Khachatryan:2015hba}. The difference with respect to the nominal results is taken as the uncertainty and is only considered for \ttbar and \tW events. This variation is correlated across years and between \ttbar and \tW events.
    \item[Top quark \pt:] Previous measurements of the differential cross section for \ttbar production have shown that the top quark has a lower average \pt value than predicted by the \POWHEG simulation~\cite{CMS:2016oae,CMS:2015rld,CMS:2015auz}. Scale factors are obtained by comparing the generated distributions of the top quark \pt with data unfolded to parton level. The difference between corrected and uncorrected shapes is taken as the uncertainty associated with the mismodelling of the top quark \pt, leaving the nominal \ttbar events untouched. This variation is correlated across years and only affects \ttbar events.
    \item[DR / DS different methods:] The difference between the DR and DS methods used to estimate the uncertainties in the \tW simulation. This variation is correlated across years.
\end{description}

\subsection{Background normalisation uncertainties}
An uncertainty of 4\%~\cite{CMS:2018fks} is taken as the uncertainty in the normalisation of the inclusive \ttbar cross section. For the \ttV, \VV, and the \nonWZ backgrounds, a normalisation uncertainty of 50\% is used, as in Ref.~\cite{Sirunyan:2018lcp}. Using the differences between the data and simulations in regions of phase space close to the signal region, a 10\% uncertainty is used for the DY background.

\section{Results}
\label{sec:results}
\subsection{Inclusive measurement}
\begin{figure}[htbp!]
    \centering
    \includegraphics[width=0.45\textwidth]{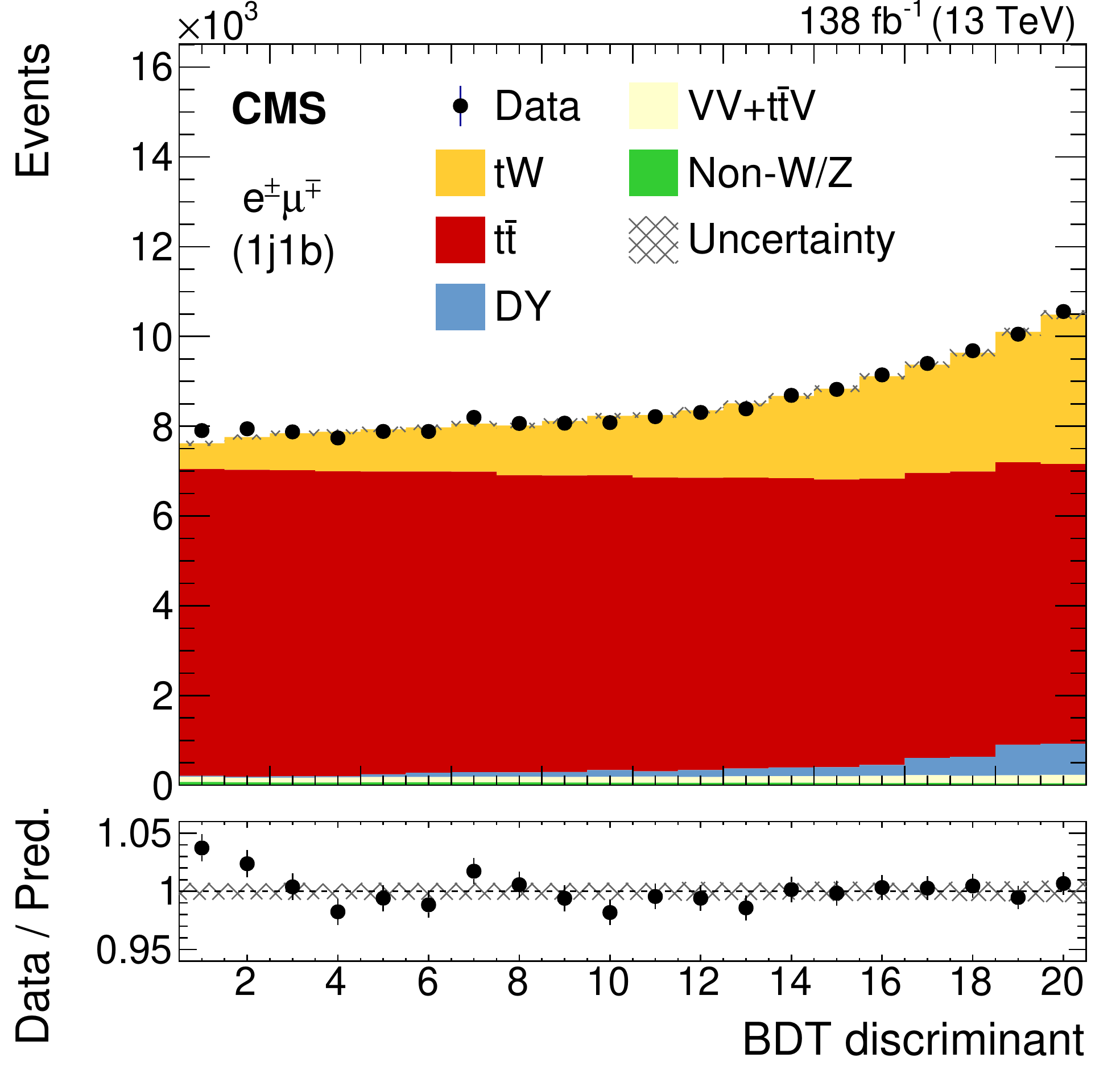}%
    \hspace*{0.05\textwidth}%
    \includegraphics[width=0.45\textwidth]{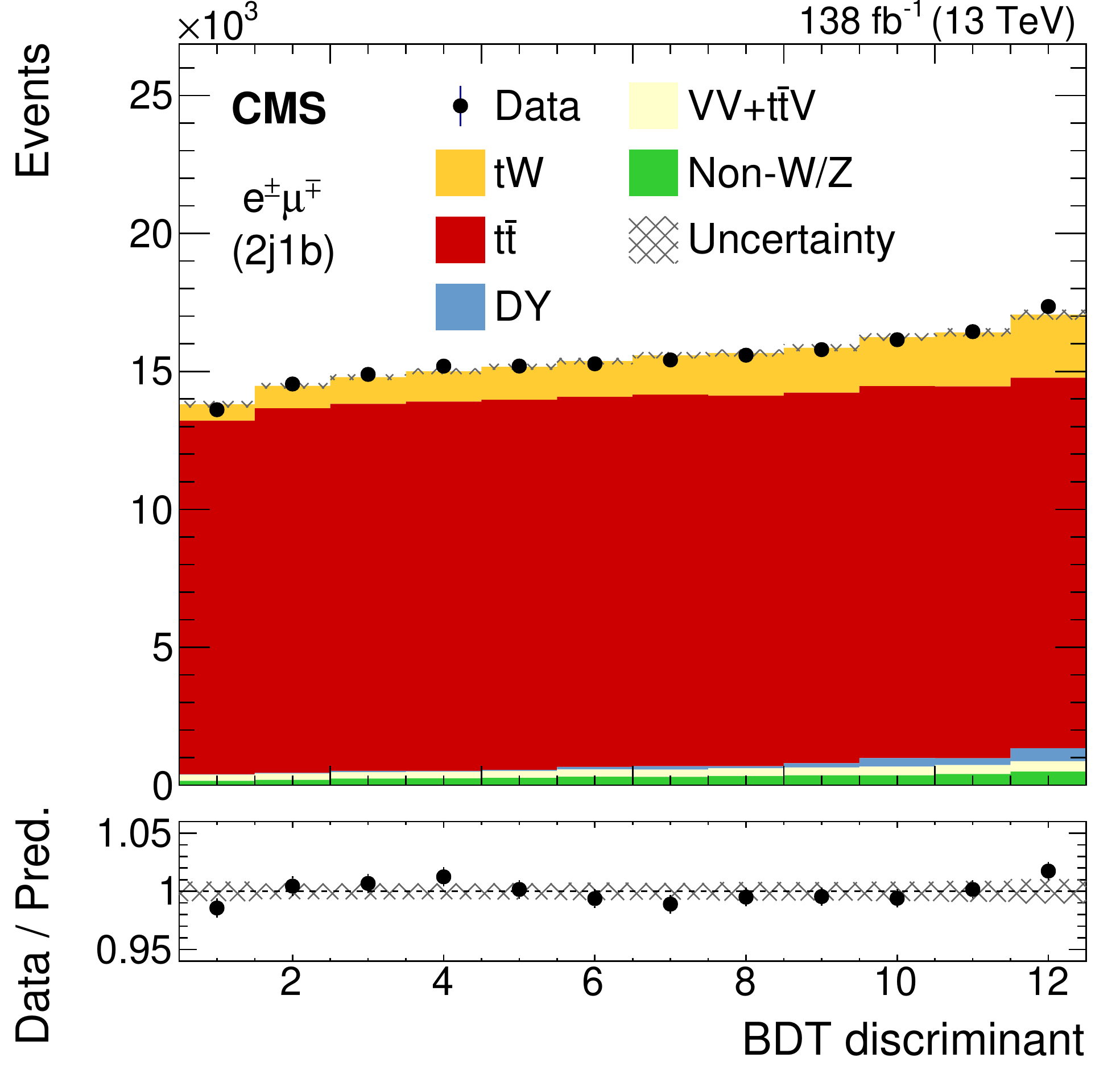} \\
    \includegraphics[width=0.45\textwidth]{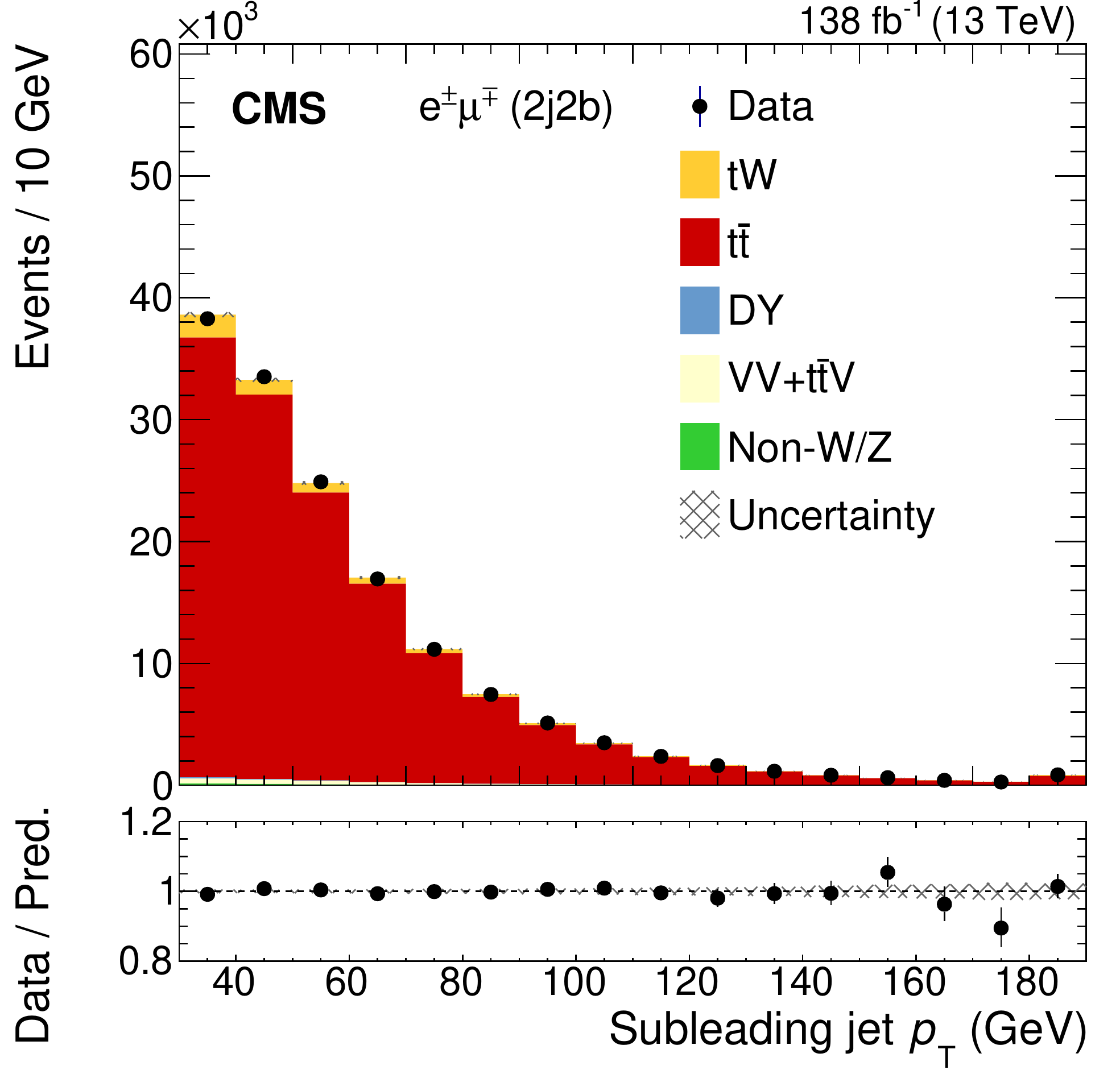}
    \caption{The distributions of the BDT outputs for events in the \njnb{1}{1} (upper left) and \njnb{2}{1} (upper right) regions, and the subleading jet \pt for the \njnb{2}{2} region (lower). The data (points) and the MC predictions (coloured histograms) after the fit are shown. The vertical bars on the points represent the statistical uncertainty in the data, and the hatched band the total uncertainty in the MC prediction. The lower panels display the ratio of the data to the sum of the MC (points) predictions after the fit, with the bands giving the corresponding uncertainties.}
    \label{fig:PostFitDistributions}
\end{figure}

The measured value for $\mu = \tWxsec/\tWxsecSM$ is obtained by maximising the likelihood function with respect to all its parameters. The fit is performed using the BDT discriminants in the \njnb{1}{1} and \njnb{2}{1} regions and the subleading jet \pt distribution in the \njnb{2}{2} region. The resulting signal strength is consistent with the SM expectations, corresponding to an inclusive cross section of
\begin{linenomath}\begin{equation}
    \tWxsec = \tWxsecobs.
\end{equation}\end{linenomath}
The distributions of the BDT discriminants in the \njnb{1}{1} and \njnb{2}{1} regions and the subleading-jet \pt distribution in the \njnb{2}{2} region after the fit are shown in Fig.~\ref{fig:PostFitDistributions}. The observed and MC predicted event yields in the three regions are given in Table~\ref{tab:postfitTable}.

\begin{table}[htbp!]
    \centering\renewcommand\arraystretch{1.1}
    \topcaption{The number of observed and MC predicted events after the fit in the \njnb{1}{1}, \njnb{2}{1}, and \njnb{2}{2} regions. The statistical uncertainties in the data and the total uncertainties in the predictions are given.}
    \begin{tabular}{lr@{${}\pm{}$}lr@{${}\pm{}$}lr@{${}\pm{}$}l}
        Process     & \multicolumn{2}{c}{\njnb{1}{1}} & \multicolumn{2}{c}{\njnb{2}{1}} & \multicolumn{2}{c}{\njnb{2}{2}}\\ \hline
        \tW         & 31\,600  & 600 & 16\,600  & 500 & 5\,500   & 200\\[\cmsTabSkip]
        \ttbar      & 131\,200 & 500 & 160\,300 & 600 & 141\,100 & 400\\
        \dy         & 3\,990   & 190 & 1\,630   & 100 & 260      & 20 \\
        \VV{+}\ttV  & 2\,800   & 300 & 3\,300   & 500 & 1\,700   & 400\\
        \nonWZ[N]   & 1\,140   & 150 & 3\,700   & 700 & 470      & 120\\[\cmsTabSkip]
        Total       & 170\,800 & 300 & 185\,400 & 400 & 149\,100 & 300\\
        Data        & 170\,900 & 400 & 185\,400 & 400 & 148\,900 & 400\\
    \end{tabular}
    \label{tab:postfitTable}
\end{table}

\begin{figure}[htbp!]
    \centering
    \includegraphics[width=\textwidth]{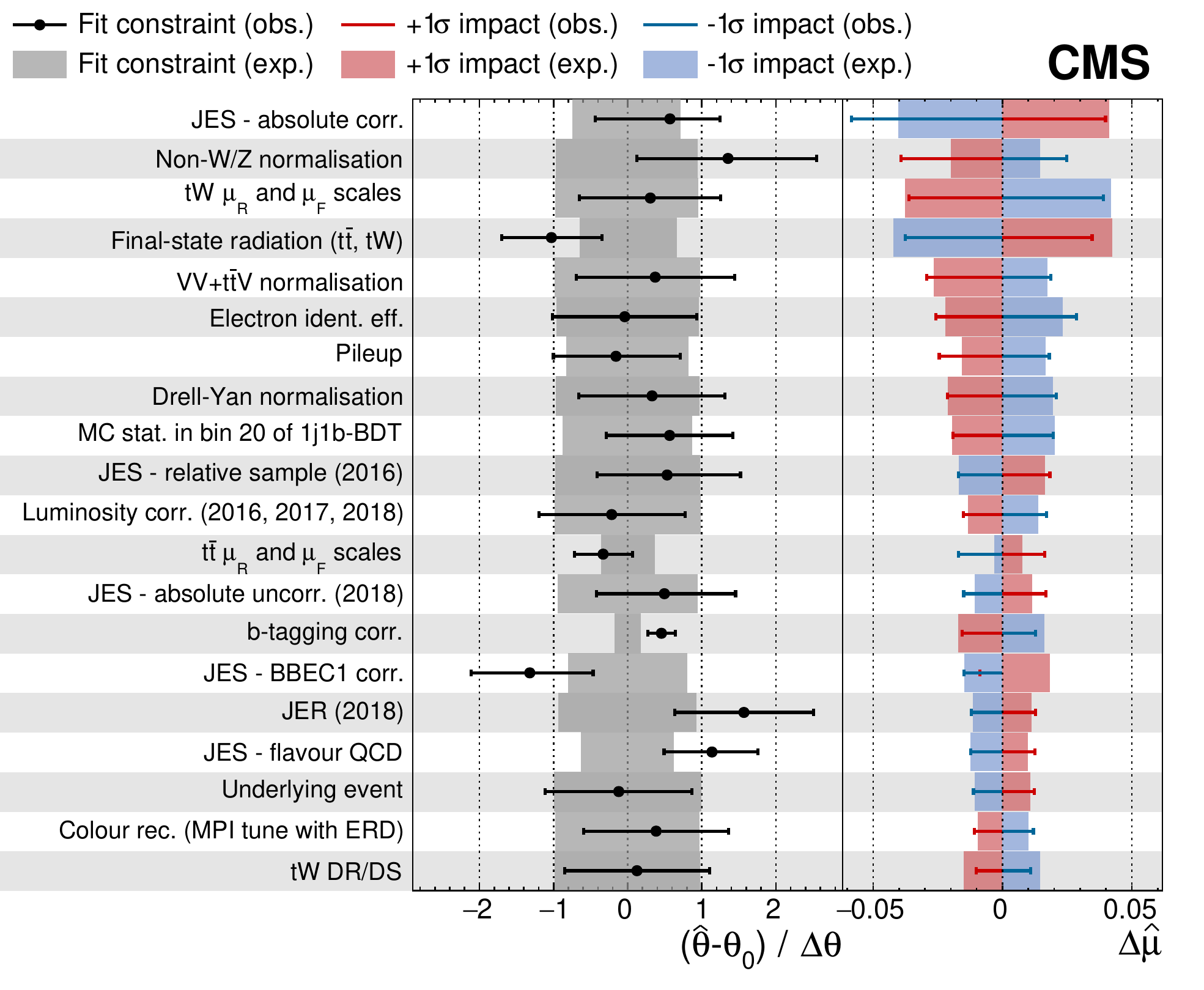}
    \caption{The 20 largest impacts \dhatmu (right column) and pulls \pull (middle column) of the nuisance parameters listed in the left column from the ML fit used to determine the inclusive \tW cross section. The horizontal bars on the pulls show the ratio of the uncertainties of the fit result to the previous ones, effectively giving the constraint on the nuisance parameter. The label ``corr.'' refers to the correlated component of the uncertainty over the three years and ``uncorr.'' the uncorrelated component for each year. The JES uncertainties are divided into several sources, where ``JES-Absolute'' groups contributions from scale corrections in the barrel, pileup corrections, and initial- and final-state radiation corrections; ``JES-Relative sample'' encodes the uncertainty in the $\eta$-dependent calibration of the jets; ``JES-BBEC1'' refers to pileup removal in the barrel (BB) and the first part of the endcaps ($1.3<\abseta<2.5$; EC1) and also a contribution from scale corrections in the barrel; and ``JES-Flavour QCD'' comes from the corrections applied to correct the different detector response to gluon and quark jets.}
    \label{fig:Impacts}
\end{figure}

The 20 largest impacts on the signal strength and the corresponding nuisance parameters are shown in Fig.~\ref{fig:Impacts}. The impact is defined as the shift \dhatmu induced in $\mu$ when the nuisance parameter $\theta$ is varied by $\pm$1 standard deviation ($\sigma$) around its best fit value. The leading uncertainties are the JES corrections, the normalisation of the \nonWZ background, the ME scales of the \tW process, and the modelling of the final-state radiation for \ttbar and \tW. Figure~\ref{fig:Impacts} also shows the pulls of the nuisance parameters, \pull, where \hattheta and \thetazero are the values after and before the fit of the nuisance parameter $\theta$, and \dtheta its uncertainty before the fit. Several nuisance parameters, such as the \PQb tagging efficiency and jet energy corrections, are significantly constrained in the fit due to their effect on the jet multiplicity. The ME scales of the \ttbar process are also constrained because of the large presence of \ttbar events in all the regions used in the fit. These constraints help in the reduction of the uncertainties during the fit, yielding the most precise measurement of the inclusive \tW cross section yet published.

\subsection{Normalised fiducial differential cross section measurements}

\begin{figure}[htbp!]
    \centering
    \includegraphics[width=0.45\textwidth]{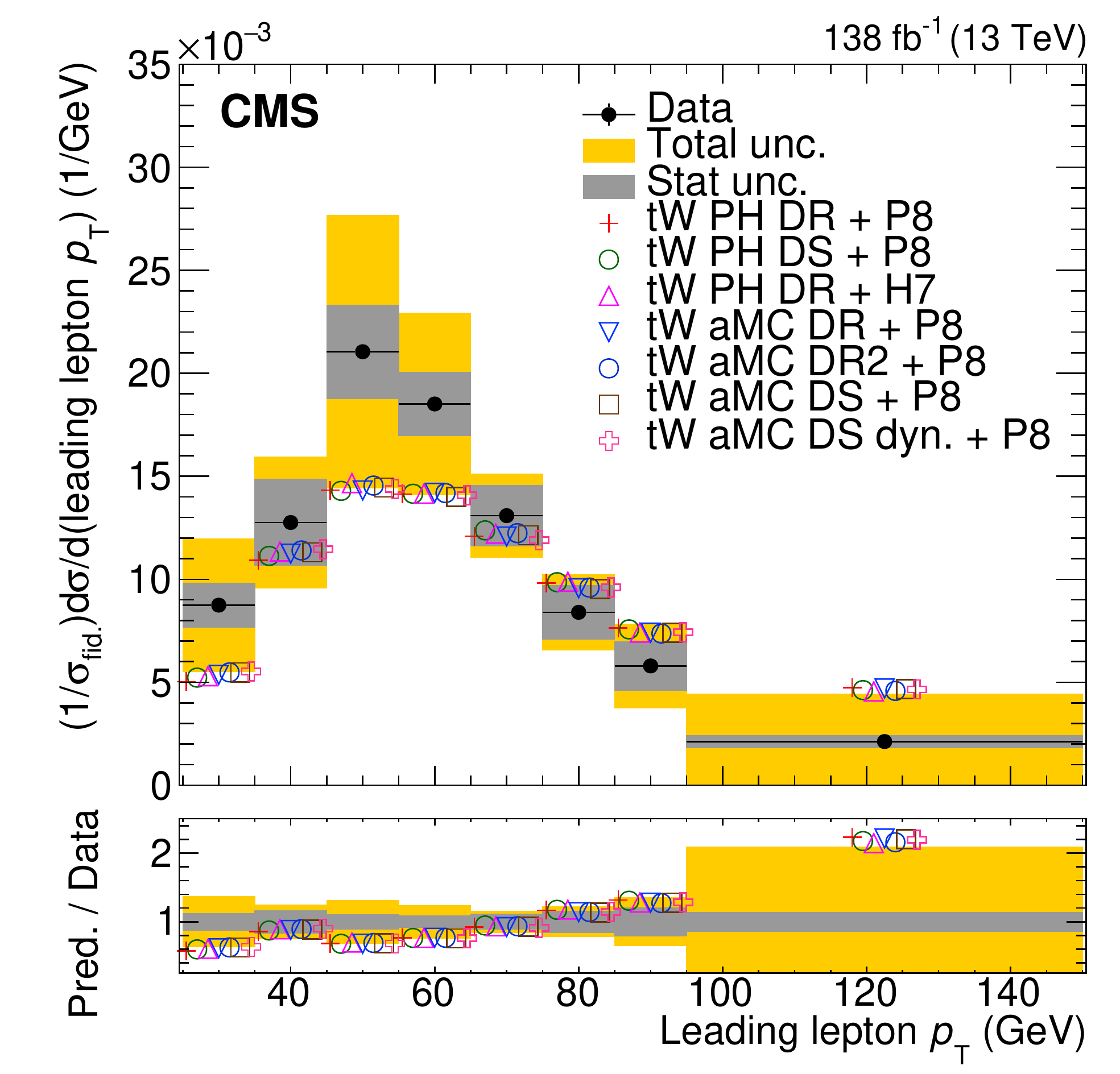}%
    \hspace*{0.05\textwidth}%
    \includegraphics[width=0.45\textwidth]{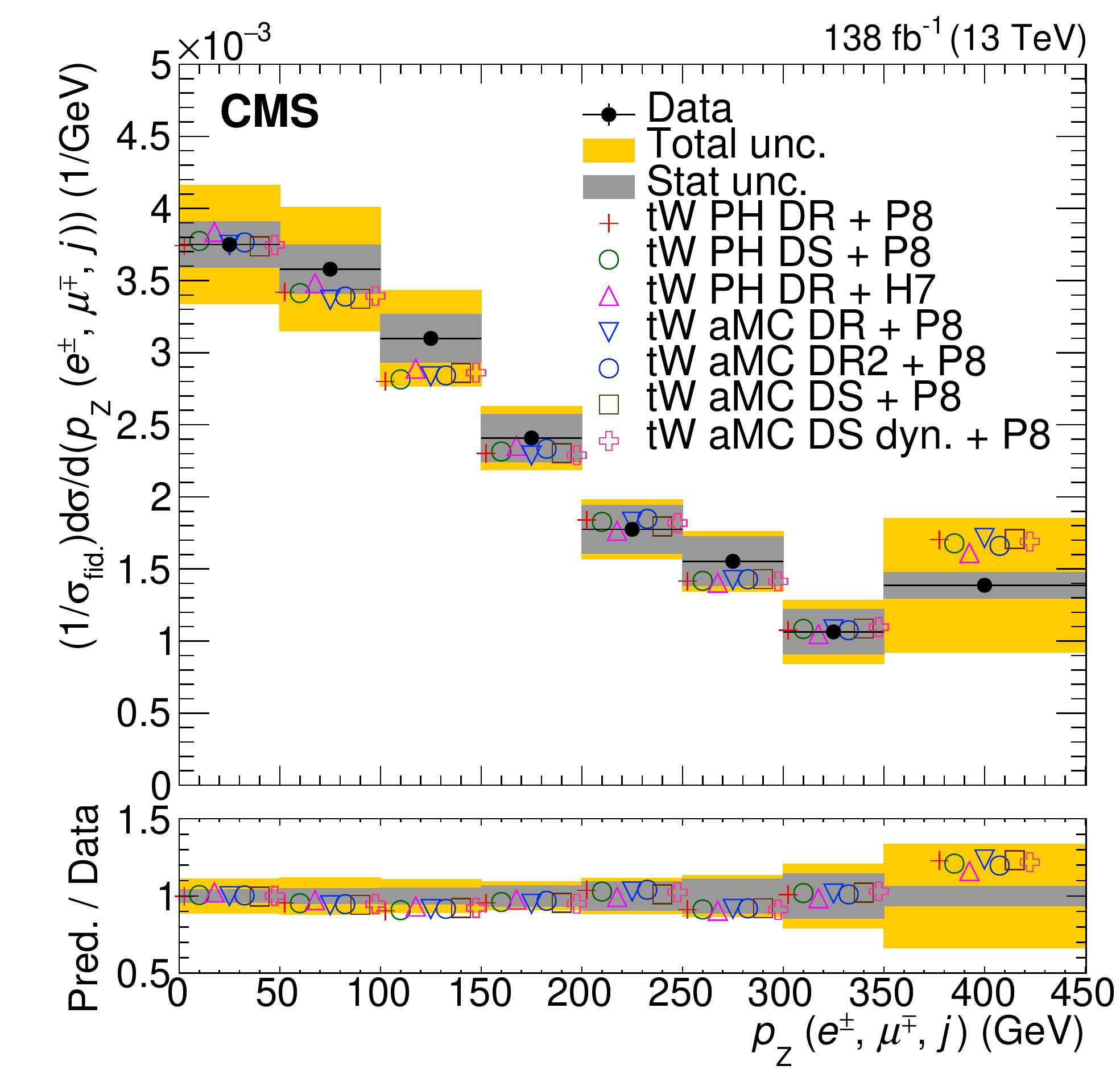} \\
    \includegraphics[width=0.45\textwidth]{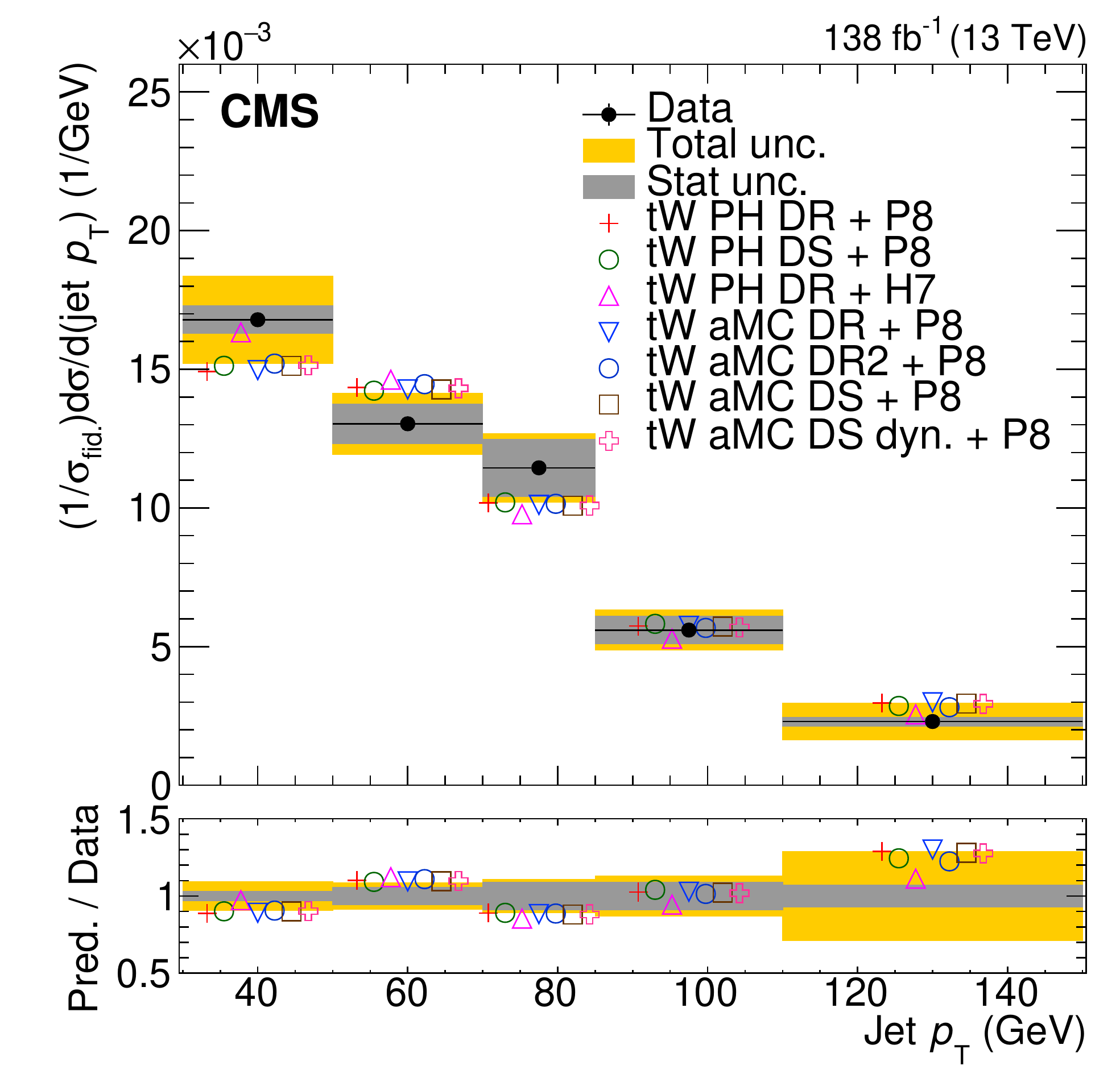}%
    \hspace*{0.05\textwidth}%
    \includegraphics[width=0.45\textwidth]{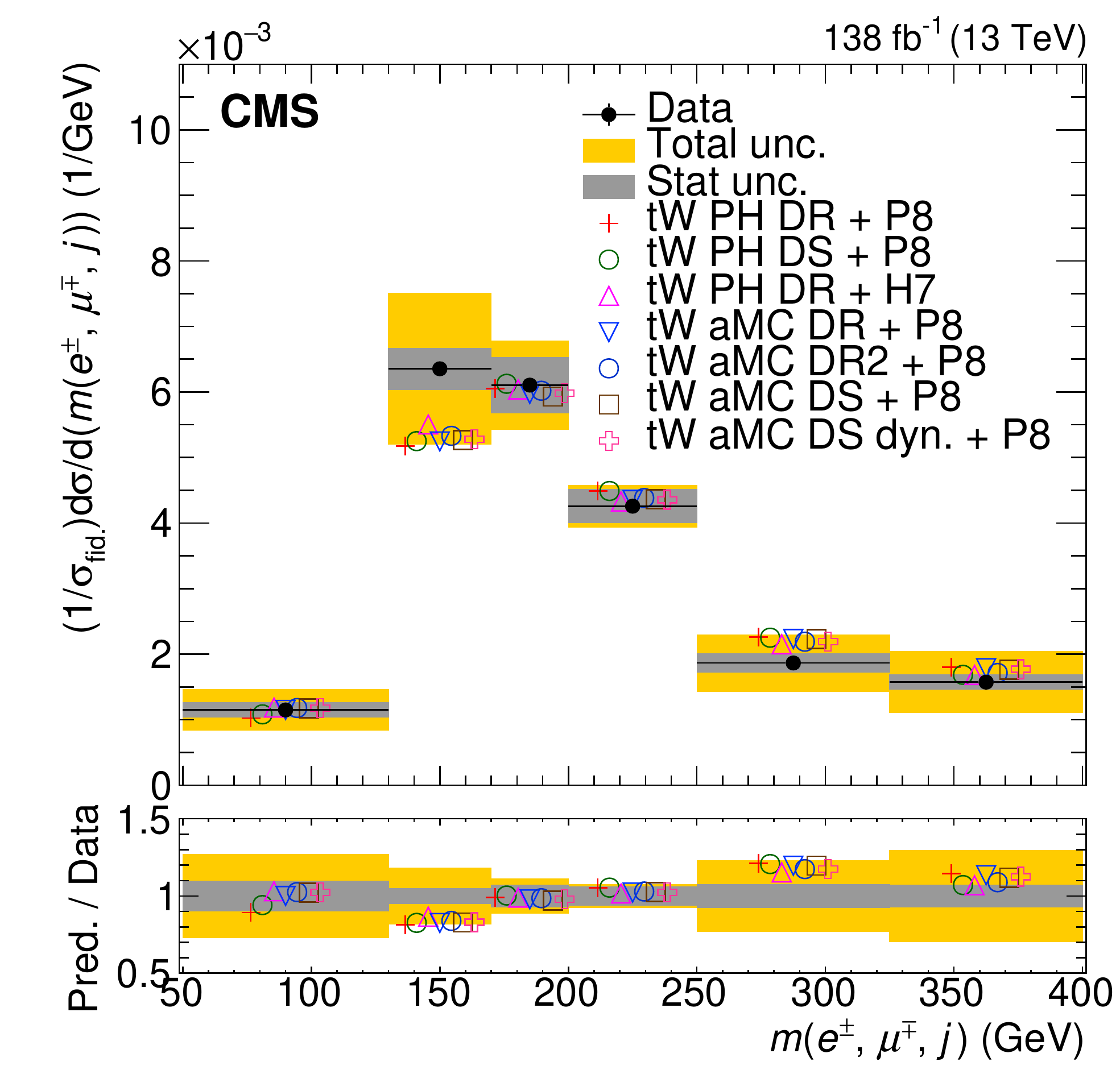} \\
    \includegraphics[width=0.45\textwidth]{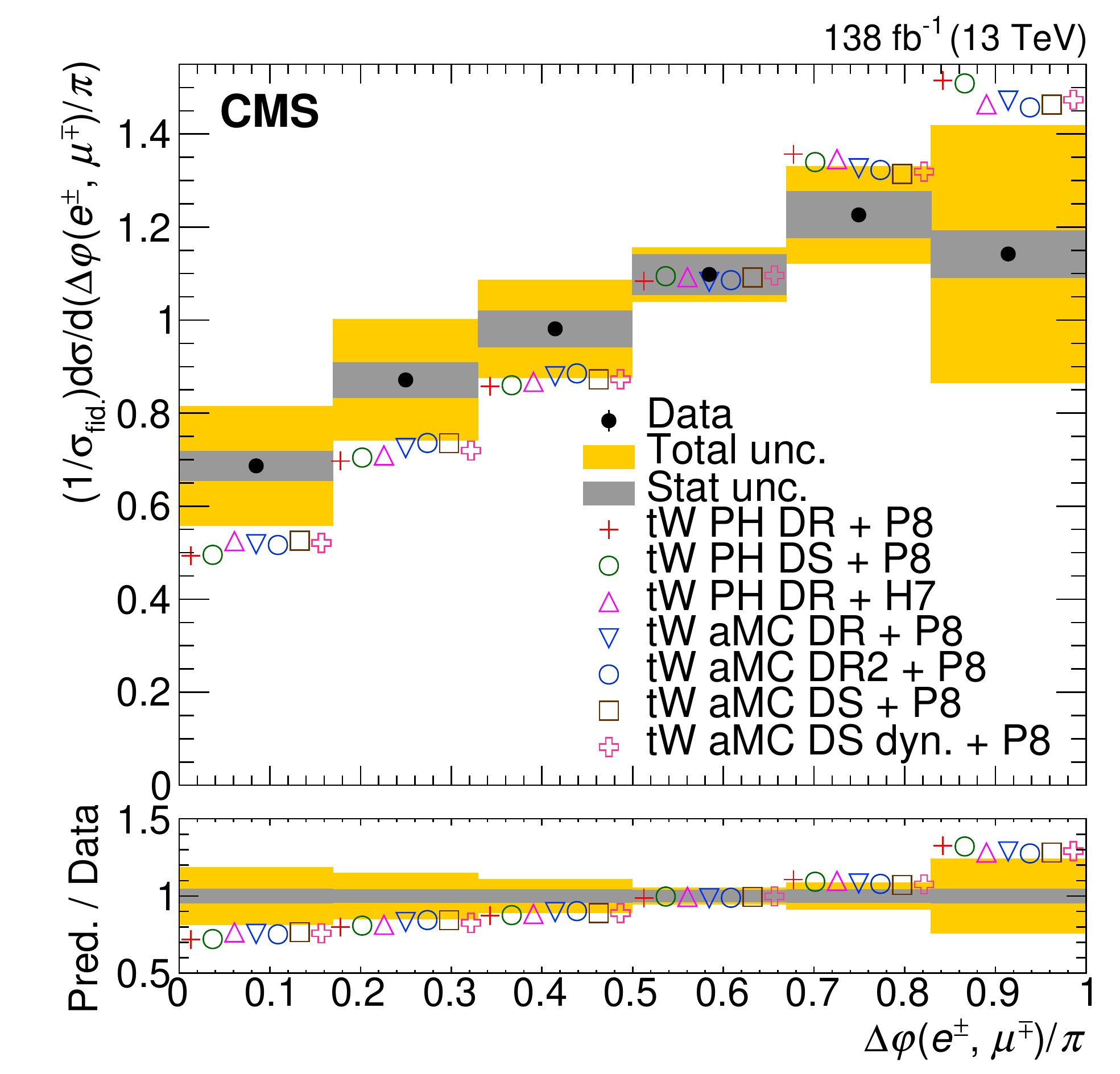}%
    \hspace*{0.05\textwidth}%
    \includegraphics[width=0.45\textwidth]{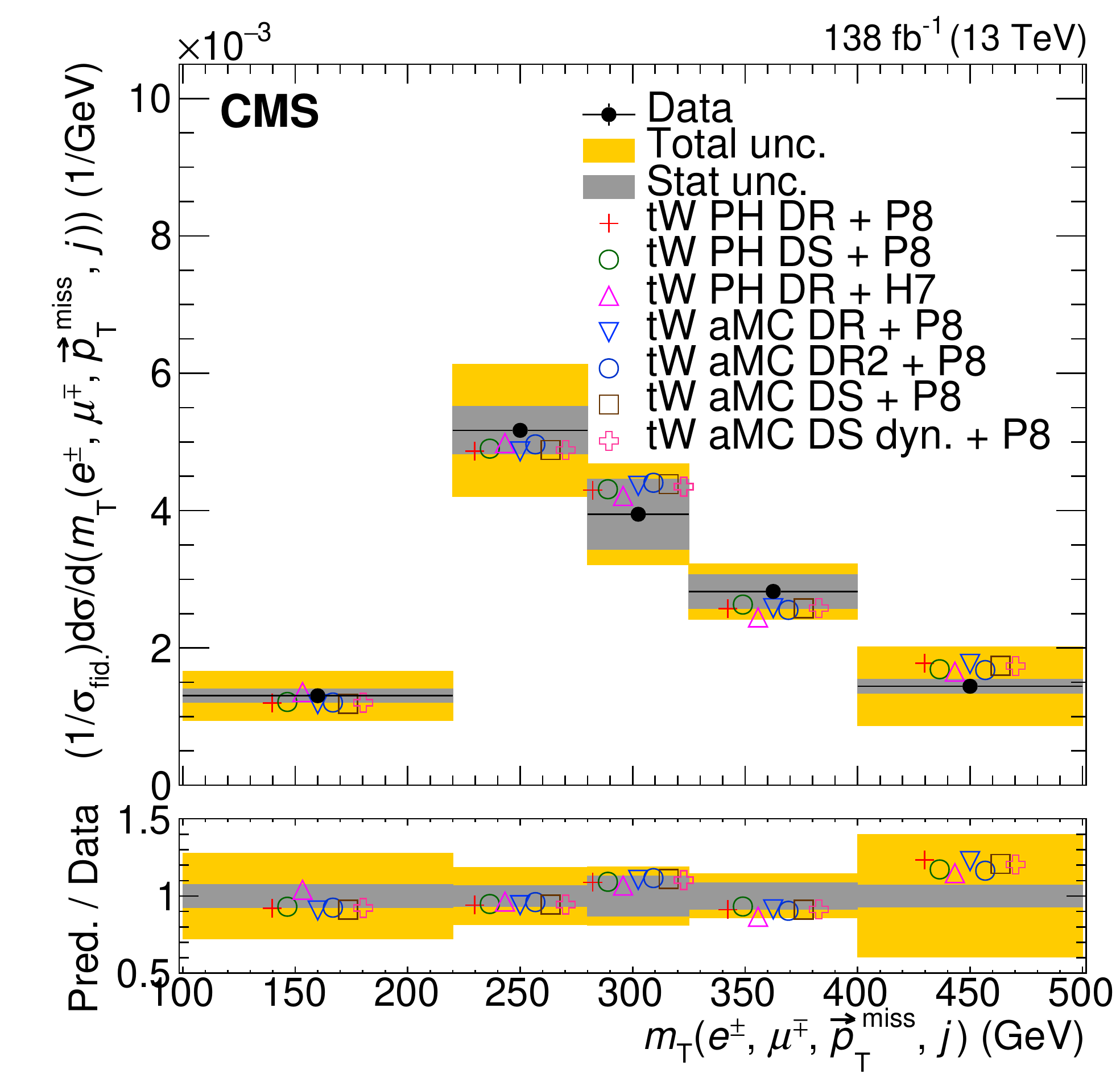}
    \caption
    {Normalised fiducial differential \tW production cross section as functions of the \pt of the leading lepton (upper left), \pzvar (upper right), \pt of the jet (middle left), \invmassvar (middle right), \deltaPhiVar (lower left), and \transmassvar (lower right). The vertical bars on the points give the statistical uncertainty in the data, the horizontal bars show the bin width. Predictions from \POWHEG (PH) + \PYTHIA8 (P8) DR and DS, \PHH (H7) DR, \MGvATNLO (aMC) + \PYTHIA8 DR, DR2, DS, and DS with a dynamic factor are also shown. The grey band represents the statistical uncertainty and the orange band the total uncertainty. In the lower panels, the ratio of the predictions to the data is shown.}
    \label{fig:particlefidbin1}
\end{figure}

\begin{table}[htbp!]
    \centering\renewcommand\arraystretch{1.1}
    \topcaption{The $p$-values from the goodness-of-fit tests comparing the six differential cross section measurements with the predictions from \POWHEG (PH) + \PYTHIA8 (P8) DR and DS and \PHH (H7) DR. The complete covariance matrix from the results and the statistical uncertainties in the predictions are taken into account.}
    \begin{tabular}{l c c c}
        Variable            &PH DR + P8&PH DS + P8&PH DR + H7\\ \hline
        Leading lepton \pt  &0.02      &0.01      &0.03\\
        Jet \pt             &0.14      &0.27      &0.01\\
        $\deltaPhiVar/\pi$  &0.26      &0.29      &0.32\\
        \pzvar              &0.70      &0.77      &0.82\\
        \transmassvar       &0.54      &0.60      &0.59\\
        \invmassvar         &0.03      &0.02      &0.28
    \end{tabular}
    \label{tab:goftests1}
\end{table}

\begin{table}[htbp!]
    \centering\renewcommand\arraystretch{1.1}
    \topcaption{The $p$-values from the goodness-of-fit tests comparing the six differential cross section measurements with the predictions from \MGvATNLO (aMC) + \PYTHIA8 DR, DR2, DS, and DS with a dynamic factor. The complete covariance matrix from the results and the statistical uncertainties in the predictions are taken into account.}
    \begin{tabular}{l c c c c}
        Variable            &aMC DR + P8&aMC DR2 + P8&aMC DS + P8&aMC DS dyn. + P8\\ \hline
        Leading lepton \pt  &0.05       &0.04        &0.03       &0.07\\
        Jet \pt             &0.15       &0.11        &0.14       &0.12\\
        $\deltaPhiVar/\pi$  &0.33       &0.40        &0.37       &0.32\\
        \pzvar              &0.76       &0.86        &0.84       &0.82\\
        \transmassvar       &0.49       &0.51        &0.48       &0.52\\
        \invmassvar         &0.09       &0.12        &0.10       &0.14
    \end{tabular}
    \label{tab:goftests2}
\end{table}

The \tW differential cross sections, normalised to the total fiducial cross section $\sigma_{\text{fid.}}$ (obtaining by summing the contents of the particle level bins), are shown in Fig.~\ref{fig:particlefidbin1} from the data and the MC predictions. Tables~\ref{tab:goftests1} and~\ref{tab:goftests2} give the $p$-values from the $\chi^2$ goodness-of-fit tests done for the six distributions, using the different MC generators and taking into account the full covariance matrix of each result, as well as the statistical uncertainties of the MC predictions. The full covariance matrix is obtained by normalising the covariance matrix extracted from the maximum likelihood fit to the measured fiducial cross section. These tests show a poorer compatibility in the leading lepton \pt, \invmassvar, and jet \pt distributions with the nominal \PHP DR prediction than in the other variables. In most of the cases, the $p$-values determined from the distributions of all the variables are similar for the other expectations. When comparing data to the predictions, there is a slight disagreement in the leading lepton \pt and the \deltaPhiVar differential cross sections. Other CMS measurements have measured similar tensions in the top quark \pt~\cite{CMS:2021vhb} and \deltaPhiVar~\cite{CMS:2019nrx} variables. All methods, DR, DR2, DS, and DS with a dynamic factor, show similar compatibility with the measurements, as well as small differences among them. This is also true for the DR predictions interfaced with \HERWIG7. The uncertainties, roughly 10--50\% in most cases, depending on the distributions and bins, are dominated by the systematic uncertainties, as in the case of the inclusive measurement.

\section{Summary}
\label{sec:summary}
Inclusive and normalised differential cross sections for the production of a top quark in association with a \PW boson are measured in proton-proton collision data at \rootsthirteen. The data, corresponding to an integrated luminosity of \lumiwounc, were recorded by the CMS detector, contain events with an electron and a muon of opposite charge.

For the inclusive measurement, the events have been categorised depending on the number of jets and jets originating from the fragmentation of bottom quarks. The signal is measured using a maximum likelihood fit to the distribution of boosted decision tree discriminants in two of the categories, and to the transverse momentum (\pt) distribution of the second-highest-\pt jet in a third category. The measured inclusive cross section is \tWxsecobs, with a total relative uncertainty of about 10\%. This is the most precise measurement of this quantity yet published. The leading uncertainty sources are the jet energy scale corrections, the normalisation in the \nonWZ background, the matrix element scales of the \tW process, and the modelling of the final-state radiation in the \ttbar and \tW processes.

The differential cross section measurements are performed as a function of six kinematical observable of the events in the fiducial phase space corresponding to the selection criteria. The results have relative uncertainties in the range of 10--50\%, depending on the measured observable, with larger values in the tails of the distributions. The uncertainties are overall systematically dominated, with the modelling sources being the dominant ones. There is overall good agreement between the measurements and the predictions from the different event generators. The different approaches used to simulate the \tW events give similar values in all the distributions, which points to small effects of  \tW/\ttbar interference on these distributions in the defined fiducial region.

\begin{acknowledgments}
We congratulate our colleagues in the CERN accelerator departments for the excellent performance of the LHC and thank the technical and administrative staffs at CERN and at other CMS institutes for their contributions to the success of the CMS effort. In addition, we gratefully acknowledge the computing centres and personnel of the Worldwide LHC Computing Grid and other centres for delivering so effectively the computing infrastructure essential to our analyses. Finally, we acknowledge the enduring support for the construction and operation of the LHC, the CMS detector, and the supporting computing infrastructure provided by the following funding agencies: BMBWF and FWF (Austria); FNRS and FWO (Belgium); CNPq, CAPES, FAPERJ, FAPERGS, and FAPESP (Brazil); MES and BNSF (Bulgaria); CERN; CAS, MoST, and NSFC (China); MINCIENCIAS (Colombia); MSES and CSF (Croatia); RIF (Cyprus); SENESCYT (Ecuador); MoER, ERC PUT and ERDF (Estonia); Academy of Finland, MEC, and HIP (Finland); CEA and CNRS/IN2P3 (France); BMBF, DFG, and HGF (Germany); GSRI (Greece); NKFIH (Hungary); DAE and DST (India); IPM (Iran); SFI (Ireland); INFN (Italy); MSIP and NRF (Republic of Korea); MES (Latvia); LAS (Lithuania); MOE and UM (Malaysia); BUAP, CINVESTAV, CONACYT, LNS, SEP, and UASLP-FAI (Mexico); MOS (Montenegro); MBIE (New Zealand); PAEC (Pakistan); MES and NSC (Poland); FCT (Portugal);MESTD (Serbia); MCIN/AEI and PCTI (Spain); MOSTR (Sri Lanka); Swiss Funding Agencies (Switzerland); MST (Taipei); MHESI and NSTDA (Thailand); TUBITAK and TENMAK (Turkey); NASU (Ukraine); STFC (United Kingdom); DOE and NSF (USA).

\hyphenation{Rachada-pisek} Individuals have received support from the Marie-Curie programme and the European Research Council and Horizon 2020 Grant, contract Nos.\ 675440, 724704, 752730, 758316, 765710, 824093, 884104, and COST Action CA16108 (European Union); the Leventis Foundation; the Alfred P.\ Sloan Foundation; the Alexander von Humboldt Foundation; the Belgian Federal Science Policy Office; the Fonds pour la Formation \`a la Recherche dans l'Industrie et dans l'Agriculture (FRIA-Belgium); the Agentschap voor Innovatie door Wetenschap en Technologie (IWT-Belgium); the F.R.S.-FNRS and FWO (Belgium) under the ``Excellence of Science -- EOS" -- be.h project n.\ 30820817; the Beijing Municipal Science \& Technology Commission, No. Z191100007219010; the Ministry of Education, Youth and Sports (MEYS) of the Czech Republic; the Hellenic Foundation for Research and Innovation (HFRI), Project Number 2288 (Greece); the Deutsche Forschungsgemeinschaft (DFG), under Germany's Excellence Strategy -- EXC 2121 ``Quantum Universe" -- 390833306, and under project number 400140256 - GRK2497; the Hungarian Academy of Sciences, the New National Excellence Program - \'UNKP, the NKFIH research grants K 124845, K 124850, K 128713, K 128786, K 129058, K 131991, K 133046, K 138136, K 143460, K 143477, 2020-2.2.1-ED-2021-00181, and TKP2021-NKTA-64 (Hungary); the Council of Science and Industrial Research, India; the Latvian Council of Science; the Ministry of Education and Science, project no. 2022/WK/14, and the National Science Center, contracts Opus 2021/41/B/ST2/01369 and 2021/43/B/ST2/01552 (Poland); the Funda\c{c}\~ao para a Ci\^encia e a Tecnologia, grant CEECIND/01334/2018 (Portugal); the National Priorities Research Program by Qatar National Research Fund; MCIN/AEI/10.13039/501100011033, ERDF ``a way of making Europe", and the Programa Estatal de Fomento de la Investigaci{\'o}n Cient{\'i}fica y T{\'e}cnica de Excelencia Mar\'{\i}a de Maeztu, grant MDM-2017-0765 and Programa Severo Ochoa del Principado de Asturias (Spain); the Chulalongkorn Academic into Its 2nd Century Project Advancement Project, and the National Science, Research and Innovation Fund via the Program Management Unit for Human Resources \& Institutional Development, Research and Innovation, grant B05F650021 (Thailand); the Kavli Foundation; the Nvidia Corporation; the SuperMicro Corporation; the Welch Foundation, contract C-1845; and the Weston Havens Foundation (USA).
\end{acknowledgments}

\bibliography{auto_generated}

\cleardoublepage \appendix\section{The CMS Collaboration \label{app:collab}}\begin{sloppypar}\hyphenpenalty=5000\widowpenalty=500\clubpenalty=5000
\cmsinstitute{Yerevan Physics Institute, Yerevan, Armenia}
{\tolerance=6000
A.~Tumasyan\cmsAuthorMark{1}\cmsorcid{0009-0000-0684-6742}
\par}
\cmsinstitute{Institut f\"{u}r Hochenergiephysik, Vienna, Austria}
{\tolerance=6000
W.~Adam\cmsorcid{0000-0001-9099-4341}, J.W.~Andrejkovic, T.~Bergauer\cmsorcid{0000-0002-5786-0293}, S.~Chatterjee\cmsorcid{0000-0003-2660-0349}, K.~Damanakis\cmsorcid{0000-0001-5389-2872}, M.~Dragicevic\cmsorcid{0000-0003-1967-6783}, A.~Escalante~Del~Valle\cmsorcid{0000-0002-9702-6359}, P.S.~Hussain\cmsorcid{0000-0002-4825-5278}, M.~Jeitler\cmsAuthorMark{2}\cmsorcid{0000-0002-5141-9560}, N.~Krammer\cmsorcid{0000-0002-0548-0985}, L.~Lechner\cmsorcid{0000-0002-3065-1141}, D.~Liko\cmsorcid{0000-0002-3380-473X}, I.~Mikulec\cmsorcid{0000-0003-0385-2746}, P.~Paulitsch, F.M.~Pitters, J.~Schieck\cmsAuthorMark{2}\cmsorcid{0000-0002-1058-8093}, R.~Sch\"{o}fbeck\cmsorcid{0000-0002-2332-8784}, D.~Schwarz\cmsorcid{0000-0002-3821-7331}, S.~Templ\cmsorcid{0000-0003-3137-5692}, W.~Waltenberger\cmsorcid{0000-0002-6215-7228}, C.-E.~Wulz\cmsAuthorMark{2}\cmsorcid{0000-0001-9226-5812}
% ...
\par}
\vskip\cmsinstskip
\dag:~Deceased\\
$^{1}$Also at Yerevan State University, Yerevan, Armenia\\
$^{2}$Also at TU Wien, Vienna, Austria\\
% ...
\end{sloppypar}
\end{document}